\documentclass[%
 reprint,
 superscriptaddress,
 amsmath,amssymb,
 aps,
]{revtex4-2}

\usepackage{graphicx}% Include figure files
\usepackage{dcolumn}% Align table columns on decimal point
\usepackage{bm}% bold math
\usepackage{hyperref}% add hypertext capabhttps://www.overleaf.com/project/5d42ee7ce928b007261846bf#cite.Berengut2018ilities

\usepackage{amssymb}
\usepackage{amsmath,bm}
\usepackage{float}
\usepackage[usenames,dvipsnames]{color}
\usepackage[export]{adjustbox}
\usepackage{gensymb}

\usepackage{soul,xcolor}

\usepackage{multirow}
\usepackage{tikz}
\usepackage{ulem}

\newcommand{\aver}[1]{\ensuremath{\langle {#1} \rangle}}

\newcommand{\drf}[0]{\ensuremath{\delta \aver{r^4}}}
\newcommand{\drt}[0]{\ensuremath{\delta \aver{r^2}}}
\newcommand{\drtsq}[0]{\ensuremath{[\drt^2]}}

\begin{document}

\title{Evidence for Nonlinear Isotope Shift in Yb$^+$ Search for New Boson
%3$\sigma$ Nonlinear Isotope Shift in Yb$^+$ Search for New Boson
}
%Measurement of 3$\sigma$ Nonlinear Isotope Shift in Yb$^+$ Search for New Boson

%\title{Precision Isotope Shift Measurement in Yb$^+$ Search for New Boson}
%\title{Nonlinear Isotope Shift in Yb$^+$ Search for Dark Matter}

\author{Ian Counts}
 \thanks{These authors contributed equally to this work.}
 \affiliation{Department of Physics and Research Laboratory of Electronics, Massachusetts Institute of Technology, Cambridge, Massachusetts 02139, USA}
\author{Joonseok Hur}
 \thanks{These authors contributed equally to this work.}
 \affiliation{Department of Physics and Research Laboratory of Electronics, Massachusetts Institute of Technology, Cambridge, Massachusetts 02139, USA}

 \author{Diana P. L. Aude Craik}
 \affiliation{Department of Physics and Research Laboratory of Electronics, Massachusetts Institute of Technology, Cambridge, Massachusetts 02139, USA}

\author{Honggi Jeon}
 \affiliation{Department of Physics and Astronomy, Seoul National University, Seoul 151-747, Korea}

 \author{Calvin Leung}
 \affiliation{Department of Physics and Research Laboratory of Electronics, Massachusetts Institute of Technology, Cambridge, Massachusetts 02139, USA}

 \author{Julian C. Berengut}
 \affiliation{School of Physics, University of New South Wales, Sydney, New South Wales 2052, Australia}
 \author{Amy Geddes}
 \affiliation{School of Physics, University of New South Wales, Sydney, New South Wales 2052, Australia}
 \author{Akio Kawasaki}
 \affiliation{W. W. Hansen Experimental Physics Laboratory and  Department of Physics, Stanford University, Stanford, California 94305, USA}

\author{Wonho Jhe}
 \affiliation{Department of Physics and Astronomy, Seoul National University, Seoul 151-747, Korea}
\author{Vladan Vuleti\'c}
 \email{vuletic@mit.edu}
 \affiliation{Department of Physics and Research Laboratory of Electronics, Massachusetts Institute of Technology, Cambridge, Massachusetts 02139, USA}

\begin{abstract}
We measure isotope shifts for five Yb$^+$ isotopes with zero nuclear spin on two narrow optical quadrupole transitions ${}^2S_{1/2} \rightarrow {}^2D_{3/2}$, ${}^2S_{1/2} \rightarrow {}^2D_{5/2}$ with an accuracy of $\sim 300$~Hz. The corresponding King plot shows a $3 \times 10^{-7}$ deviation from linearity at the 3 $\sigma$ uncertainty level. Such a nonlinearity can indicate physics beyond the Standard Model (SM) in the form of a new bosonic force carrier, or arise from higher-order nuclear effects within the SM. We identify the quadratic field shift as a possible nuclear contributor to the nonlinearity at the observed scale, and show how the nonlinearity pattern can be used in future, more accurate measurements to separate a new-boson signal from nuclear effects.

%While a nonlinearity in the King plot can in principle indicate physics beyond the Standard Model (SM), our precision atomic-structure calculations indicate that similarly sized deviations may arise from isotopic shape change $\delta \aver{r^4}$ of the nuclear charge distribution. In the future, precision measurements on other narrow optical transition available for Yb$^+$ and Yb can be used to distinguish nonlinearities within the SM from possible non-SM effects.

%Achievements: measurement of delta r2 in combination with atomic calculations, in combination with existing absolute measurement this determines optical transition frequencies of all stable bosonic isotopes with xxx precision. 

\end{abstract}

\maketitle

%%%%% ---- Main text start ---- %%%%

The Standard Model (SM) of particle physics describes virtually all measurements of elementary particles exquisitely well, and yet various indirect evidence points to physics beyond the SM. This evidence includes the preponderance of dark matter of unknown composition in our Universe, astronomically observed with several different methodologies such as the rotation curves of galaxies \cite{GallaxyRotation}, the motion of colliding galaxy clusters \cite{GallaxyCollision}, gravitational lensing \cite{GravLensing}, and the power spectrum of the cosmic microwave background \cite{CMB}. Physics beyond the SM is also being probed in various laboratory experiments, such as high-energy collisions \cite{PDG}, searches for weakly interacting massive particles \cite{PDG}, axions, and axionlike particles \cite{ALPs}, precision measurements of the electric dipole moments of elementary particles \cite{EDM}, and other precision tests \cite{BSMwAMO}.

\begin{figure}
    \centering
    \includegraphics[width=\columnwidth,trim=170 165 13 145,clip]{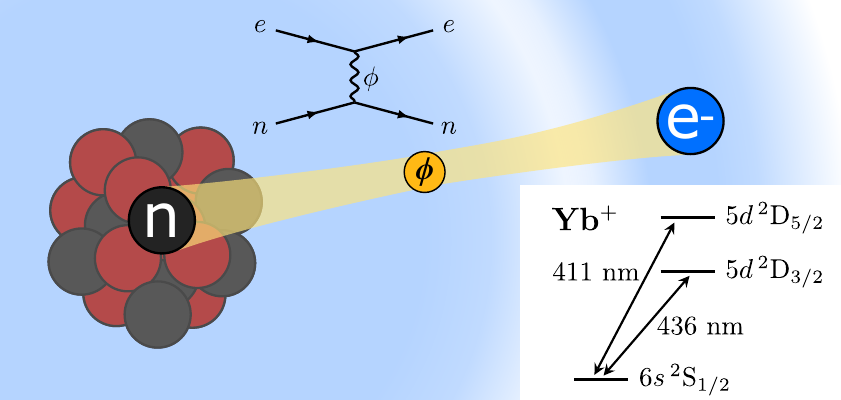}
    \caption{New intra-atomic force between electron ($e^-$) and neutron ($n$) mediated by the virtual exchange of a hypothetical new boson $\phi$. The coupling results in a Yukawa-like potential that modifies the atomic energy levels and can be probed with isotope-shift spectroscopy. We perform precision measurements of the long-lived states ${}^{2}D_{3/2}$, $^{2}D_{5/2}$ on individual trapped Yb$^+$ ions.}
    \label{fig:n-e_interaction}
\end{figure}

Dark-matter candidates can be characterized by their mass, spin, and interactions. In the intermediate mass range from $\sim 100$~eV$/c^2$ to $\sim 100$~MeV$/c^2$, a new method has been proposed to search for a dark-matter boson $\phi$ that couples to quarks and leptons \cite{Delaunay2017,Berengut2018}. The virtual exchange of $\phi$ between neutrons and electrons in an atom would result in a Yukawa-like potential in addition to the Coulomb potential of the nucleus (see Fig. \ref{fig:n-e_interaction}). The corresponding shift in energy levels and transition frequencies is too small to be detected by directly comparing spectroscopic data to (much less accurate) atomic-structure calculations, but could potentially be detected through precision isotope-shift measurements \cite{Gebert2015,Manovitz2019,Knollmann2019,Miyake2019} that allow one to sidestep electronic-structure calculations. In particular, the scaled isotope shifts on two different transitions exhibit a linear relationship (King plot \cite{King1984}), and Refs. \cite{Delaunay2017,Berengut2018} argue that a deviation from linearity can indicate a new force mediator $\phi$. Such studies are particularly timely as recent experiments analyzing nuclear decay in {}$^8$Be and {}$^4$He have observed a 7 $\sigma$ deviation from the SM \cite{Krasznahorkay2018,Krasznahorkay2016,Krasznahorkay2019new} that could be potentially explained by a new boson with a mass of 17~MeV/$c^2$ ($X17$ boson) \cite{Feng2016,Feng2017,Jentschura2018,NA64_2018}. According to Ref. \cite{Berengut2018}, measurements of optical transitions with a resolution of 1~Hz in select atomic systems could probe this scenario. However, higher-order effects within the SM can result in nonlinearities that limit the sensitivity to new physics \cite{Flambaum2018,Mikami2017,Tanaka2019,Allehabi2020}.
%Presently, only order-of-magnitude estimations of such nonlinearities have been provided \cite{Flambaum2018,Mikami2017,Tanaka2019}.

In this Letter, we report a precision measurement of the isotope shift for five isotopes of Yb$^+$ ions with zero nuclear spin on two narrow optical quadrupole transitions (${}^2S_{1/2} \rightarrow {}^2D_{3/2}, {}^2D_{5/2}$) with an accuracy of $\sim 300$~Hz. Displaying the data in a King plot \cite{King1984}, we observe a deviation from linearity at the $10^{-7}$ level, corresponding to 3 standard deviations $\sigma$. With four independent isotope-shift data points available, we further introduce a novel parametrization of the nonlinearity pattern that can be used to distinguish between nonlinearities of the same magnitude but different physical origin. At the current level of precision, the observed nonlinearity pattern is consistent with both a new boson, and the quadratic field shift (QFS) \cite{Flambaum2018} that we identify as the leading source of nonlinearity within the SM by means of precision electronic-structure calculations. In the future, more accurate measurements on the present and other optical transitions in Yb and Yb$^+$ \cite{Huntemann2016,Barber2006,Safronova2018} can discriminate between effects within and outside the SM.

\begin{table*}
    \caption{\label{table:IsotopeShift}Inverse-mass differences $\mu_{ji}$ and measured isotope shifts $\nu_{ji}$ between next-neighboring pairs of five even Yb$^+$ isotopes. $\mu_{ji}$ is calculated from the mass of Yb$^+$ ions with the ionization energy set to 6.254~eV \cite{AME2016_1,AME2016_2,Rana2012}. The nuclear size difference $\drt$ is deduced from $\nu_{ji}$ using the calculated parameters $F_{\alpha}^{\mathrm{CI}}=-15.852$~GHz/fm$^2$, $F_{\beta}^{\mathrm{CI}} = -16.094$~GHz/fm$^2$, $F_{\alpha}^{\mathrm{MBPT}}=-16.570$~GHz/fm$^2$, $F_{\beta}^{\mathrm{MBPT}} = -16.771$~GHz/fm$^2$, $K_{\alpha}^{\mathrm{CI}} = -1678.3~\mathrm{GHz\cdot u}$, and $K_{\beta}^{\mathrm{CI}} = -1638.5~\mathrm{GHz\cdot u}$ (see the Supplemental Material \cite{SM}). The uncertainties given here and throughout the paper for $\nu_{\alpha ji}$ and $\nu_{\beta ji}$ indicate 1$\sigma$ statistical uncertainties; the estimated systematic uncertainties on these quantities are $<20$\% of the statistical uncertainties (see the Supplemental Material \cite{SM}). The (170,174) pair is directly measured as a cross-check [the measurements (170,174) and (170,172), (172,174) agree within 2 $\sigma$] and to improve precision (see the Supplemental Material \cite{SM}). In the calculations of $\drt_{ji}$ from the measured isotope shifts, the average of the values for $\alpha$ and $\beta$ is given (the difference between transitions is less than 0.2\%) (see the Supplemental Material \cite{SM}), and the values of $K_\alpha$ and $K_\beta$ from the CI calculations are used for both CI and MBPT. For the data from Ref. \cite{Angeli2013} (last column), only the statistical errors are presented in the parentheses, while the systematic errors from the calculation of the electronic factors are much larger.
    }
	\begin{ruledtabular}
		\begin{tabular}{ccccccc}
		Isotope pair & \multirow{2}{*}{$\mu_{ji}$ ($10^{-6} \mathrm{u}^{-1}$)}  & $\nu_{\alpha ji}$ (kHz) & $\nu_{\beta ji}$ (kHz) & \multicolumn{3}{c}{$\drt_{ji}$ (fm$^2$)} \\ $(j, i)$ & & $\alpha: {}^2S_{1/2} \rightarrow {}^2D_{5/2}$ & $\beta: {}^2S_{1/2} \rightarrow {}^2D_{3/2}$ & CI & MBPT & Reference \cite{Angeli2013} \\
            \hline
			(168, 170) & 70.113 698(46)\quad\ \   & 2 179 098.93(21) & 2 212 391.85(37) & -0.156 & -0.149 & -0.1561(3)  \\
			(170, 172) & 68.506 890 50(63) & 2 044 854.78(34) & 2 076 421.58(39) & -0.146 & -0.140 & -0.1479(1) \\
			(172, 174) & 66.958 651 95(64) & 1 583 068.42(36) & 1 609 181.47(22) & -0.115 & -0.110 & -0.1207(1) \\
			(174, 176) & 65.474 078 21(65) & 1 509 055.29(28) & 1 534 144.06(24) & -0.110 & -0.105 & -0.1159(1) \\
			\hline
			(170, 174) &  & 3 627 922.95(50) & 3 685 601.95(33) & & &
		\end{tabular}
	\end{ruledtabular}
    
\end{table*}

\begin{figure}
   \centering
    \includegraphics[width=\columnwidth]{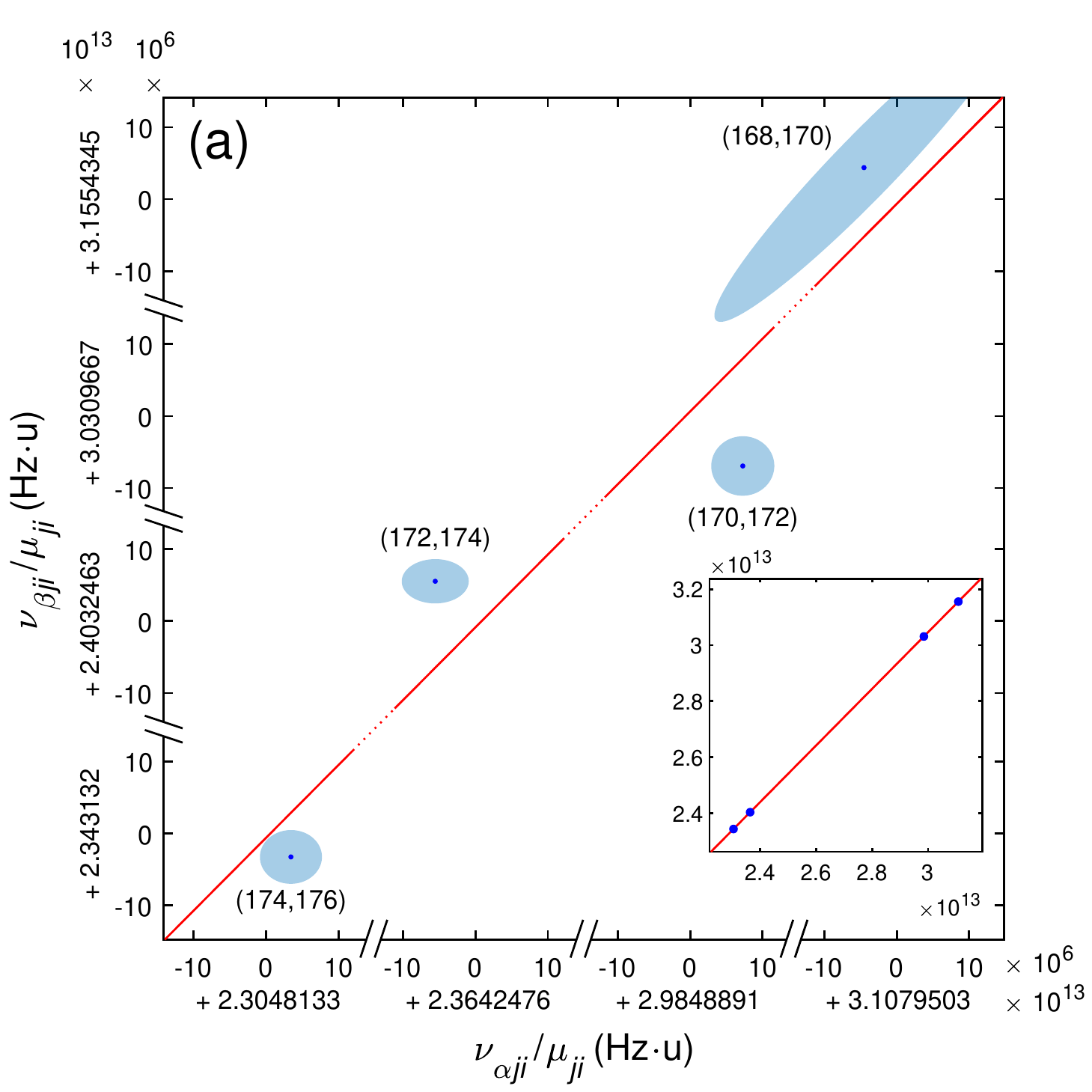}
    \includegraphics[width=\columnwidth]{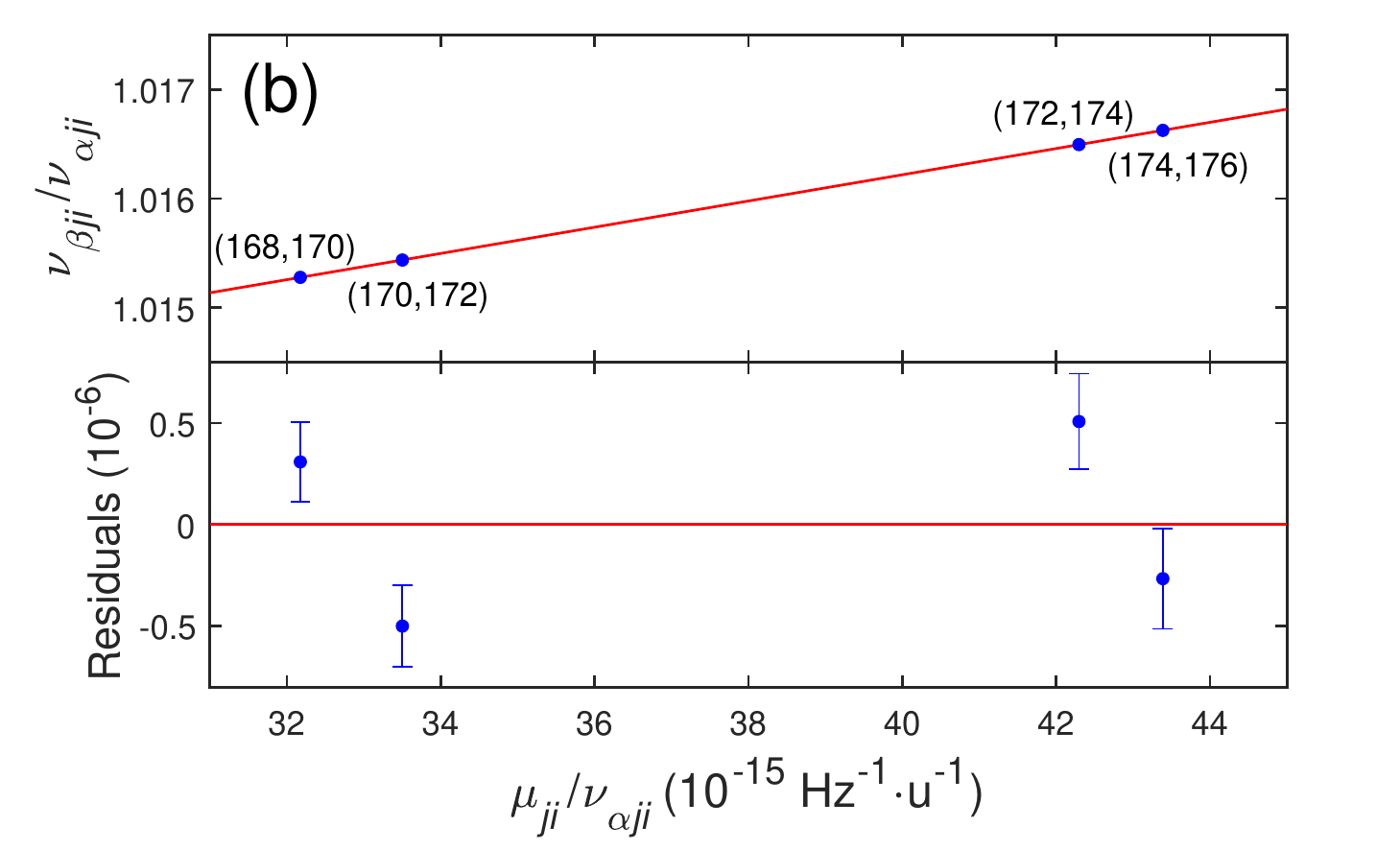}
    \caption{(a) Standard King plot [Eq.~(\ref{Eq:KingPlot})] for the $\alpha=411$~nm, ${}^2S_{1/2} \rightarrow {}^2D_{5/2}$, and $\beta =436$~nm, ${}^2S_{1/2} \rightarrow {}^2D_{3/2}$ transitions for pairs of neighboring even Yb$+$ isotopes. The inset shows the full King plot. The main figure is zoomed into the data points by a factor of $10^6$. A deviation from linearity (red line) by 3 standard deviations $\sigma$ is observed. The larger diagonal uncertainty for the $(168,170)$ pair is due to the larger mass uncertainty for the ${}^{168}$Yb$^+$ isotope \cite{AME2016_1,AME2016_2,Rana2012} (see the Supplemental Material \cite{SM}). (b) Frequency-normalized King plot [Eq.~(\ref{Eq:ModifiedKingPlot})] and residuals. The error bars and error ellipses indicate $1 \sigma$.} 
    \label{fig:king_plot}
\end{figure}

Our measurements are performed with individual ${}^j\text{Yb}^+$ ions ($j \in \{168,170,172,174,176 \}$) trapped in a linear Paul trap, and Doppler cooled on the $6s\,{}^2S_{1/2} \rightarrow 6p\,{}^2P_{1/2}$ transition to typically 500~$\mu$K \cite{Cetina2013}. We perform optical precision spectroscopy on the transitions to two long-lived excited states (with electron configurations $[\text{Xe}]4f^{14}6s\,{}^2S_{1/2} \rightarrow [\text{Xe}]4f^{14}5d\,{}^2D_{3/2}, {}^2D_{5/2}$) using light at the wavelengths 411 and 436~nm, respectively. The probe light is generated by a frequency-doubled Ti:Sapphire laser that is frequency stabilized to an ultralow-thermal-expansion cavity, achieving a short-term stability of $\sim 200$~Hz. Typically, 1~mW of 411-nm light (0.2~mW of 436-nm light) is focused to a waist of $w_0 = 60~\mathrm{\mu m}$  ($w_0 = 15~\mathrm{\mu m}$) at the location of the ion (see Supplemental Material (the Supplemental Material) \cite{SM} for details).

Coherent optical Ramsey spectroscopy is carried out with two $\frac{\pi}{2}$ pulses of 411- or 436-nm light, lasting 5~$\mu$s each, separated by 10~$\mu$s. This is followed by readout of the state, performed using an electron-shelving scheme \cite{Taylor1997} (see the Supplemental Material \cite{SM}). A small magnetic field of typically $\sim 1.1$~G is applied to separate the different Zeeman components of the $S \rightarrow D$ transition. Frequency scans are taken over the central Ramsey fringes of the two symmetric Zeeman components with the lowest magnetic-field sensitivity to find the center frequency of the transition (see the Supplemental Material \cite{SM}).

The measurement on one isotope is averaged typically for 30 minutes before we switch to a next-neighboring isotope by adjusting various loading, cooling, and repumper laser frequencies. We typically perform three interleaved measurements of each isotope to determine an isotope shift, allowing us to reach a precision on the order of $\sim 300$~Hz (see Table~\ref{table:IsotopeShift} and Fig. \ref{fig:king_plot}), limited mainly by drifts in the frequency stabilization of the probe laser to the ultrastable cavity (see the Supplemental Material \cite{SM}). 

The frequency shift $\nu_{\alpha ji}$ between isotope {}$^j$Yb and {}$^i$Yb on an optical transition $\alpha$ can be written as a sum of terms that factorize into a nuclear part (with subscript $ji$) and an electronic part (with subscript $\alpha$) \cite{King1984,Mikami2017,Delaunay2017}
\begin{equation}
\nu_{\alpha ji} = F_\alpha \drt_{ji} + K_\alpha \mu_{ji} + G_\alpha \drtsq_{ji} + \upsilon_{ne} D_\alpha a_{ji}
\label{Eq:IsotopeShift}
\end{equation}
Here $\delta \aver{r^2}_{ji} \equiv \aver{r^2}_j - \aver{r^2}_i$ is the difference in squared charge radii $r$ between isotope $j$ and $i$, $\mu_{ji} \equiv 1/m_j - 1/m_i$ is the inverse-mass difference, $\drtsq_{ji} \equiv (\drt_{jl})^2 - (\drt_{il})^2$ for some fixed isotope $l$ (the choice of $l$ is irrelevant to the nonlinearity) (see the Supplemental Material \cite{SM}), and $a_{ji} = j-i$ is the difference in neutron number. The quantity $\upsilon_{ne}=(-1)^{s+1} y_n y_e /(4 \pi \hbar c)$ is the product of the coupling factors of the new boson to the neutron $y_n$ and electron $y_e$, creating a Yukawa-like potential given by $V_{ne}(r) = \hbar c \, \upsilon_{ne} \exp(-r/\lambdabar_c)/r$ for a boson with spin $s$, mass $m_\phi$, and reduced Compton wavelength $\lambdabar_c=\hbar/(m_\phi c)$ \cite{Mikami2017,Delaunay2017}.

For heavy elements like Yb, the first term in Eq.~(\ref{Eq:IsotopeShift}) associated with the change in nuclear size $ \drt$ [``field shift" (FS)] dominates, while the second term is due to the electron's reduced mass and momentum correlations between electrons (``mass shift"). According to our electronic-structure calculations (see below), the third (QFS) term associated with the square of nuclear size $\drtsq_{ji}$ represents the leading-order nonlinearity \cite{Mikami2017,Flambaum2018} within the SM for Yb. The last term describes the isotope shift due to the Yukawa-like potential associated with the new boson $\phi$. The quantities $F, K, G, D$ are determined by the electronic wave functions of the transition \cite{Delaunay2017,Berengut2018,Mikami2017}; see the Supplemental Material \cite{SM}. Note that the effect of the next-leading order Seltzer moment \cite{Seltzer1969,Mikami2017} associated with $\drf$ is absorbed into the QFS term; see the Supplemental Material \cite{SM}.

The first two terms in Eq.~(\ref{Eq:IsotopeShift}) lead to a linear relationship between the isotope shifts (King plot \cite{King1984}) when one considers two different transitions $\alpha, \beta$,
\begin{equation}
\overline{\overline{\nu}}_{\beta ji} = K_{\beta \alpha} + F_{\beta \alpha} \overline{\overline{\nu}}_{\alpha ji} + G_{\beta \alpha} \overline{\overline{\drtsq}}_{ji} + \upsilon_{ne} D_{\beta \alpha} \overline{\overline{a}}_{ji}
\label{Eq:KingPlot}
\end{equation}
Here we define $F_{\beta \alpha} \equiv F_{\beta}/F_{\alpha}$, $P_{\beta \alpha} \equiv P_\beta - F_{\beta \alpha} P_\alpha$ for $P \in \{K, G, D \}$, while $\overline{\overline{z}}_{ji} \equiv z_{ji}/ \mu_{ji}$ for $z \in \{\nu_{\alpha},\nu_{\beta, }, \drtsq, a \}$, is the inverse-mass-normalized quantity. For our purposes, where the FS dominates, the influence of mass and frequency errors is more transparent if we instead write a modified linear relationship for the frequency-normalized quantities $\overline{x}_{ji} \equiv x_{ji}/\nu_{\alpha ji}$ for $x \in \{\nu_{\beta}, \mu,  \drtsq, a \}$
\begin{equation}
\overline{\nu}_{\beta ji} = F_{\beta \alpha} + K_{\beta \alpha} \overline{\mu}_{ji} + G_{\beta \alpha} \overline{\drtsq}_{ji} + \upsilon_{ne} D_
{\beta \alpha} \overline{a}_{ji}
\label{Eq:ModifiedKingPlot}
\end{equation}
To analyze the experimental results in this work, the transitions and isotopes are assigned as follows: $\alpha = {}^2S_{1/2} \rightarrow {}^2D_{5/2}$ (411~nm), $\beta = {}^2S_{1/2} \rightarrow {}^2D_{3/2}$ (436~nm), $j \in \{168,170,172,174\}$ with $i=j+2$, and $l = 172$.

The inset in Fig. \ref{fig:king_plot}(a) confirms the general linear relationship for the inverse-mass-normalized isotope shifts in a standard King plot corresponding to Eq.~(\ref{Eq:KingPlot}) for the two transitions $\alpha$ and $\beta$. However, when we zoom in by a factor of $10^6$ [main figure \ref{fig:king_plot}(a)], we observe a small deviation from linearity, in the range 0.5 -- 1~kHz in frequency units for a given data point. The frequency-normalized King plot associated with Eq.~(\ref{Eq:ModifiedKingPlot}), as displayed in Fig. \ref{fig:king_plot}(b), illustrates that due to the smallness of the slope, i.e. the mass-shift electronic factor $K_{\beta \alpha}$, the mass error along the horizontal axis $\overline{\mu}_{ji}$ has a negligible effect. For all points taken together, the nonlinearity is nonzero at the level of 3 $\sigma$ (see the Supplemental Material \cite{SM}).

\begin{figure}
    \centering
    \includegraphics[width=\columnwidth,trim=50 10 30 10,clip]{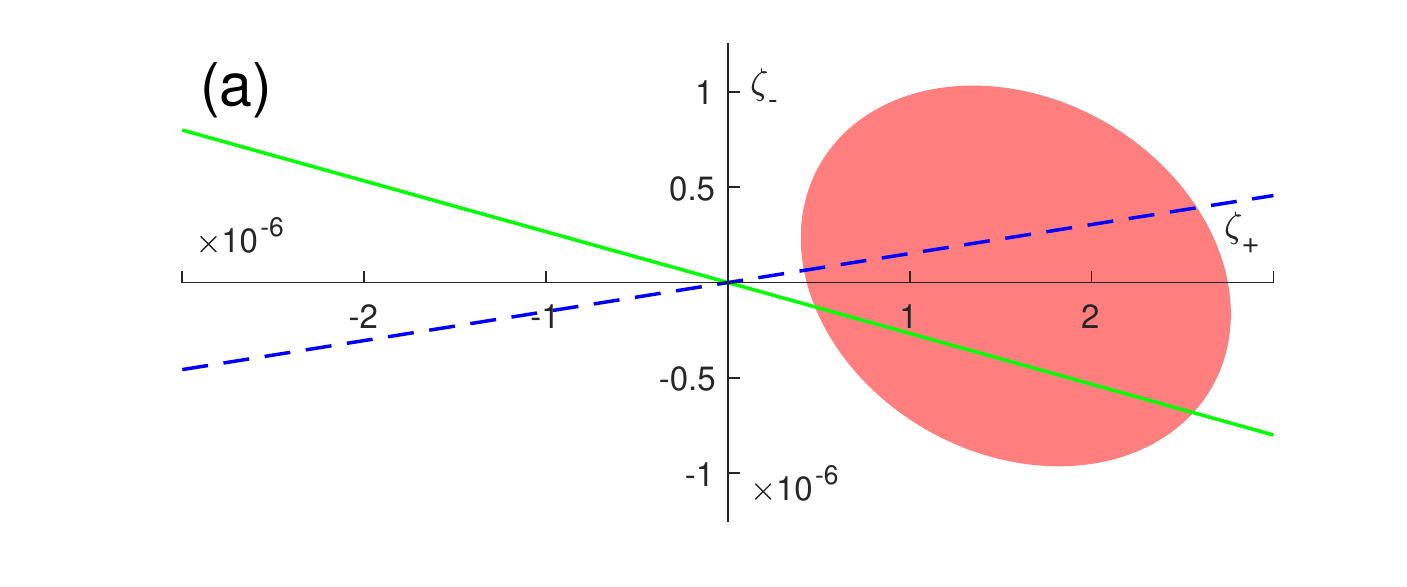}
    \includegraphics[width=\columnwidth,trim=50 0 30 10,clip]{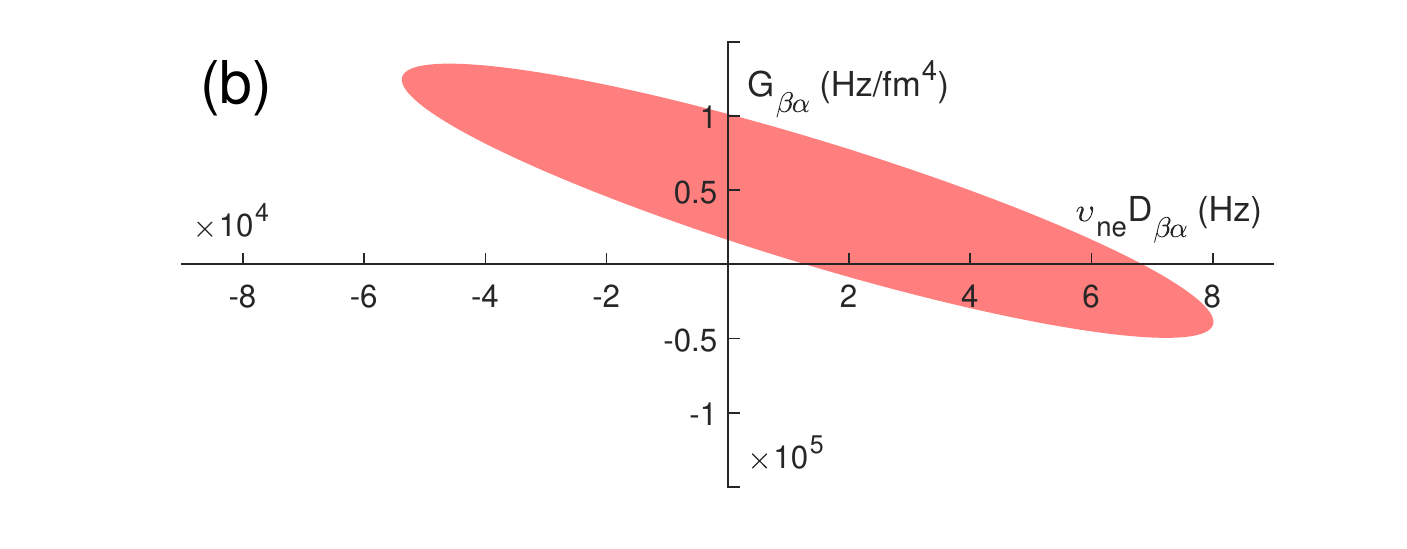}
    \caption{(a) Nonlinearity measure $(\zeta_+,\zeta_-)$ for next-neighbor isotope pairs. The red shaded region indicates the 95\% confidence interval from our data. The green solid line and the blue dashed line indicate the required ratio $\zeta_-/\zeta_+$ if the nonlinearity is purely due to a new boson $\phi$ and the QFS, respectively. (b) Experimental nonlinearity measure along the axes of a new boson ($x$-axis) and the QFS ($y$-axis).}
    \label{fig:zeta_map}
\end{figure}

With four independent isotope pairs, we can quantify not only the magnitude of the nonlinearity, but also an associated pattern further characterizing the nonlinearity. To this end, we introduce two dimensionless nonlinearity measures
\begin{equation}
    \zeta_{\pm} \equiv d_{168}-d_{170} \pm (d_{172} -d_{174})
\end{equation}
where $d_{j} \equiv \overline{\nu}_{\beta ji} - f(\overline{\mu}_{ji})$ are the vertical deviations of the four data points $\overline{\nu}_{\beta ji}$ in Fig. \ref{fig:king_plot}(b) from the linear fit $f$. $\zeta_+$ and $\zeta_-$ characterize the two possible nonlinearities for four data points, a zigzag shape with deviation pattern $+ - + -$, and a curved nonlinearity with deviation pattern $+--+$, respectively. Any given nonlinearity can be represented by a point in the $\zeta_+ \zeta_-$ plane [see Fig. \ref{fig:zeta_map}(a)]. A nonlinearity that arises from the coupling of the $\phi$ boson to the neutron number corresponds to a fixed nonlinearity pattern, and hence a given line through the origin (see the Supplemental Material \cite{SM}). The same argument holds for the QFS. Our observed nonlinearity lies close to both lines representing pure coupling to a new boson and the QFS, respectively. The experimental uncertainty region in Fig. \ref{fig:zeta_map}(a) can be decomposed into its possible QFS and new-boson components, as shown in Fig. \ref{fig:zeta_map}(b). It highlights the relative contributions of the two sources of nonlinearity, ranging from pure new boson to pure QFS contribution  at the current level of uncertainty. With increased measurement precision, it will be possible to separate the two contributions.

In order to convert the observed nonlinearity, as represented by $\zeta_{\pm}$, into a physical quantity such as the coupling $\upsilon_{ne}$, we need to determine the associated electronic wave functions. To cross-check our numerical simulations for systematic errors, we use two different methods, the Dirac-Hartree-Fock method \cite{Grant1980,Dyall1989} followed by the configuration interaction (CI) method \cite{Jonsson1996,Porsev2009,Fawcett1991,Biemont1998}, using the software package GRASP2018 \cite{FroeseFischer2018}, and many-body perturbation theory (MBPT) \cite{Dzuba1985} implemented in \textsc{amb}{\footnotesize i}\textsc{t} \cite{Kahl2019}. We calculate $F_{\beta\alpha}^{\mathrm{CI}} = 1.0153$ and $F_{\beta\alpha}^{\mathrm{MBPT}} = 1.0121$, within 0.2\% and 0.07\% of our experimental value $F_{\beta\alpha}^{\mathrm{exp}} = 1.01141024(86)$, respectively. For the mass shift, that is more difficult to calculate accurately, we find $K_{\beta\alpha}^{\mathrm{CI}} = 65~\mathrm{GHz \cdot u}$ (see the Supplemental Material \cite{SM}), within a factor of 2
from the experimental value $K_{\beta\alpha}^{\mathrm{exp}} = 120.208(23)~\mathrm{GHz\cdot u}$. The calculated wave functions in combination with the measured frequency shift can also be used to extract the nuclear size difference $\drt$ (see the Supplemental Material \cite{SM}), in good agreement with other results \cite{Angeli2013}; see Table \ref{table:IsotopeShift}. We also calculate $G_{\beta \alpha}^{\mathrm{CI}} = 232$~kHz/fm$^4$ and $G_{\beta \alpha}^{\mathrm{MBPT}} = -36$~kHz/fm$^4$ for the QFS, indicating a large systematic uncertainty in the calculation of this small term. The experimentally constrained range in Fig. \ref{fig:zeta_map}(b) (24 -- 94~kHz/fm$^4$) (see the Supplemental Material \cite{SM}) lies between the two calculated values.

\begin{figure}
    \centering
    \includegraphics[width=\columnwidth,trim=0 0 45 0,clip]{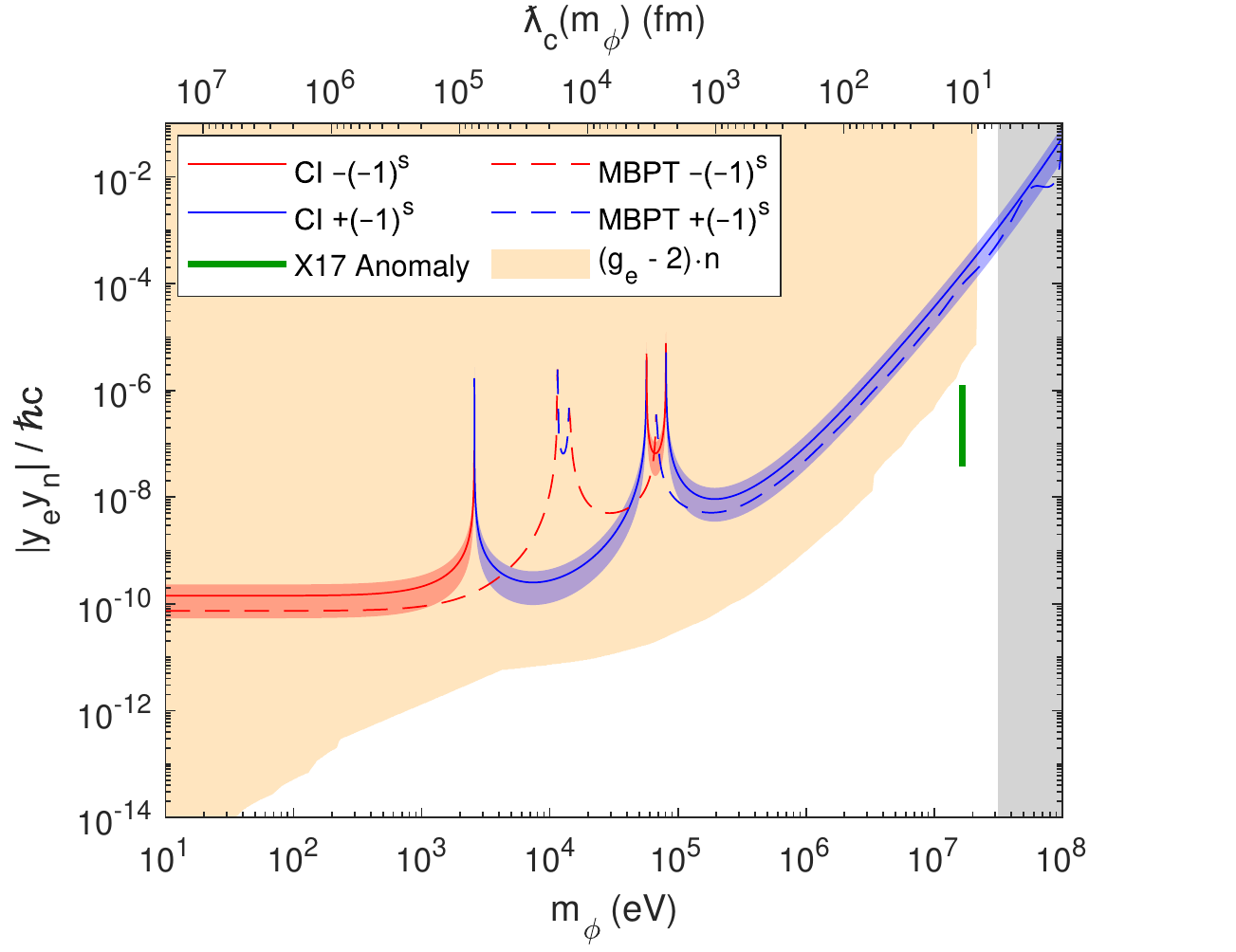}
    \caption{\label{fig:yeyn_vs_m} Product of couplings $\left|y_e y_n\right|$ of a new boson vs. boson mass $m_\phi$ (bottom) and reduced Compton wavelength (top), plotted under the assumption that the observed nonlinearity in Fig. \ref{fig:zeta_map} is dominated by the new boson. The solid line is for the CI calculation, and the dashed line is for the MBPT calculation. If the nonlinearity has a contribution from the QFS, then $\left|y_e y_n\right|$ lies below this line. The gray shade indicates the region inside the nucleus. The sign of $y_e y_n$ is color-coded: red for $-(-1)^{s}$ and blue for $+(-1)^{s}$ for a spin-$s$ boson. The 95\% confidence intervals from the statistical uncertainty in the measured isotope shift are shown as shaded areas along the solid line. The systematic uncertainty due to the wave function calculation is much larger, especially in the high-mass region. The thick green line indicates the preferred coupling range for the $X17$ boson from the Be/He anomaly \cite{Krasznahorkay2018,Krasznahorkay2016,Krasznahorkay2019new,Feng2016,Feng2017,Jentschura2018}. The yellow shaded area shows the constraint from electron $g_e-2$ measurements \cite{Hanneke2008,Hanneke2011,Aoyama2012,Bouchendira2011,Davoudiasl2014} combined with neutron scattering measurements \cite{Barbieri1975,Leeb1992,Pokotilovski2006,Nesvizhevsky2008} (from Ref. \cite{Berengut2018}).}
\end{figure}

Using the electronic-structure calculations, we can determine a boundary on the new-boson coupling from our data. Figure \ref{fig:yeyn_vs_m} shows the upper bound on the product of couplings $|y_e y_n|$. It is obtained by dividing the experimental value of $\upsilon_{ne} D_{\beta\alpha}$ from Fig.~\ref{fig:zeta_map}(b) (determined with the assumption that the effect of the new boson dominates the nonlinearity; i.e., $G_{\beta \alpha}=0$), by $(-1)^{s+1} D_{\beta\alpha}(m_\phi)/(4\pi\hbar c)$ from the atomic-structure calculations (see the Supplemental Material \cite{SM} for the calculation of $D_{\beta\alpha}$). The calculations with the CI and the MBPT methods agree with each other to better than a factor of 2 over most of the mass range $m_\phi$. The upper bound from our data on $|y_e y_n|$ is $\sim 200$ times larger than the preferred coupling range for the $X17$ boson \cite{Feng2016,Feng2017}, and 2 orders of magnitude larger than the bound estimated in Ref. \cite{Berengut2018} from the combination of $g-2$ measurements on the electron and neutron scattering data. We note, however, that the limit on $|y_e|$ depends on additional assumptions about the new boson's spin and the symmetries of the interaction.

%due to the near-cancellation of $G_\beta / G_\alpha$ and $F_\beta / F_\alpha$.

Finally, since the absolute optical frequency of the ${}^2S_{1/2} \rightarrow {}^2D_{5/2}$ transition for $^{172}$Yb$^+$ has recently been measured with precision at the Hz level \cite{fuerst2020}, the absolute frequencies for all the other bosonic isotopes can be deduced from our isotope shift measurements. The results are summarized in Table \ref{tab:AbsoluteFreq411}.

\begin{table}
    \caption{\label{tab:AbsoluteFreq411} Frequencies of the ${}^2S_{1/2} \rightarrow {}^2D_{5/2}$ transition. %The uncertainties listed here (except for those for $^{172}$Yb$^+$) are dominated by the uncertainties in the measured isotope shifts.} The (170, 174) pair is used to improve the precision of the measured isotope shifts, see SMat \cite{SM}.
    }
	\begin{ruledtabular}
	            \begin{tabular}{cll}
            Isotope & Absolute frequency (kHz) & Ref. \\
            \hline
            168 & 729 481 090 980.86(36) & This work \\
            170 & 729 478 911 881.93(30) & This work \\
            172 & 729 476 867 027.2068(44) & \cite{fuerst2020} \\
            174 & 729 475 283 958.85(31) & This work \\
            176 & 729 473 774 903.56(42) & This work \\
        \end{tabular}
    \end{ruledtabular}
\end{table}

In the future, the measurement precision can be increased by several orders of magnitude by cotrapping two isotopes \cite{Manovitz2019,Knollmann2019}. This improvement, also in combination with measurements on additional transitions, such as the ${}^2S_{1/2} \rightarrow {}^2F_{7/2}$ octupole transition in Yb$^+$ \cite{Roberts1997} or clock transitions in neutral Yb \cite{Barber2006,Safronova2018}, will allow one to discriminate between nonlinearities of different origin. Characterizing the nonlinearities arising from within the SM can provide new information about the nucleus \cite{Reinhard2020}, especially in combination with improved electronic-structure calculations. On the other hand, if evidence for a new boson should emerge from the improved measurements, it can be independently verified by performing similar measurements on other atomic species \cite{Berengut2018}, such as Ca/Ca$^+$ \cite{Solaro2020,Knollmann2019,Mortensen2004}, Sr/Sr$^+$ \cite{Miyake2019,Manovitz2019}, Nd$^+$ \cite{Bhatt2020}, or on highly charged ions \cite{Micke2020,Yerokhin2020,Silwal2018,Kozlov2018}, as well as on molecules like Sr$_{2}$ \cite{McGuyer2015}. Unstable isotopes (e.g. ${}^{166}$Yb with a half-life of $\sim 2.4$~days) can be used to increase the number of points in the King Plot, providing strong further constraints on the origin of the nonlinearity. The generalization of nonlinearity measures $\zeta_\pm$ for more isotopes or transitions is discussed in the Supplemental Material \cite{SM}.

This work was supported by the NSF and NSF CUA. This project received funding from the European Union's Horizon 2020 research and innovation programme under the Marie Sklodowska-Curie Grant No. 795121. C.L. was supported by the U. S. Department of Defense (DoD) through the National Defense Science \& Engineering Graduate Fellowship (NDSEG) Program. J.C.B. is supported by the Australian Research Council (DP190100974). A.K. acknowledges the partial support of a William M. and Jane D. Fairbank Postdoctoral Fellowship of Stanford University. We thank M. Drewsen, V. V. Flambaum,  P. Harris, N. Huntemann, T. Mehlst\"{a}ubler, R. Milner, R. Ozeri, E. Peik, G. Perez, R.F. Garcia Ruiz, Y. Soreq, J. Thaler, and T. Zelevinsky for interesting discussions, W. Nazarewicz and P. Reinhard for providing information about the Yb nucleus, and V. Dzuba for pointing out the nuclear quadrupole deformation as a potential significant source of nonlinearity \cite{Allehabi2020}.

%\nocite{CetinaThesis2011,Olmschenk2007,Martensson-Pendrill1994,Feldker2018,Sugiyama2000,Gerz1988,Efron1993,Lizuain2007,Berkeland1998,Safronova2012,Roy2017,Nandy2014,Dube2013,Degenhardt2005,Sanner2018,TrappedChargedParticlesBook,Dzuba1996,Dzuba2010,Safronova2009,Ekman2019,Blundell1987,Roberts1999,Tamm2009,Webster2010,Fricke2004,Papoulia2016,Puchalski2010,Palmer1987,DeVries1987,Sasanuma1979}

%\bibliography{IsotopeShiftPaperRefs}
%% \bibliographystyle{bst/apsrev4-2}
%\bibliographystyle{bst/apsrev4-0_edited}

%%%%% ---- End of main text ---- %%%%%%

\providecommand{\noopsort}[1]{}\providecommand{\singleletter}[1]{#1}%

\end{document}

\title{Supplemental Material:\\ 
		Evidence for Nonlinear Isotope Shift in Yb$^+$ Search for New Boson}
	
	\author{Ian Counts}
	\thanks{These authors contributed equally to this work.}
	\affiliation{Department of Physics and Research Laboratory of Electronics, Massachusetts Institute of Technology, Cambridge, Massachusetts 02139, USA}
	\author{Joonseok Hur}
	\thanks{These authors contributed equally to this work.}
	\affiliation{Department of Physics and Research Laboratory of Electronics, Massachusetts Institute of Technology, Cambridge, Massachusetts 02139, USA}
	
	\author{Diana P. L. Aude Craik}
	\affiliation{Department of Physics and Research Laboratory of Electronics, Massachusetts Institute of Technology, Cambridge, Massachusetts 02139, USA}
	
	\author{Honggi Jeon}
	\affiliation{Department of Physics and Astronomy, Seoul National University, Seoul 151-747, Korea}
	\author{Calvin Leung}
	\affiliation{Department of Physics and Research Laboratory of Electronics, Massachusetts Institute of Technology, Cambridge, Massachusetts 02139, USA}
	
	\author{Julian C. Berengut}
	\affiliation{School of Physics, University of New South Wales, Sydney, New South Wales 2052, Australia}
	\author{Amy Geddes}
	\affiliation{School of Physics, University of New South Wales, Sydney, New South Wales 2052, Australia}
	\author{Akio Kawasaki}
	\affiliation{W. W. Hansen Experimental Physics Laboratory and  Department of Physics, Stanford University, Stanford, California 94305, USA}
	
	\author{Wonho Jhe}
	\affiliation{Department of Physics and Astronomy, Seoul National University, Seoul 151-747, Korea}
	\author{Vladan Vuleti\'c}
	\email{vuletic@mit.edu}
	\affiliation{Department of Physics and Research Laboratory of Electronics, Massachusetts Institute of Technology, Cambridge, Massachusetts 02139, USA}

	\maketitle
	
	%%%%% ---- Main text start ---- %%%%
	\beginsupplement
	
	\section{Experimental Details}
	We trap a single Yb$^+$ ion 135$\,\mu$m above the surface of a lithographic microchip, described in detail in Ref.~\cite{CetinaThesis2011}. The ion is Doppler-cooled on the $6s\,{}^{2}S_{1/2} \rightarrow 6p\,{}^{2}P_{1/2}$ transition by a 369-nm laser beam aligned parallel to the chip surface, and at a diagonal to to the trap axis. This beam has a component along all motional modes of the trapped ion, and hence cools all motional degrees of freedom simultaneously. The ion occasionally decays to the $5d\, {}^{2}D_{3/2}$ state from $^{2}P_{1/2}$ during the cooling cycle (branching ratio = 0.5\% \cite{Olmschenk2007}; occurs once every $\sim$150 $\mu$s in our system), and is subsequently returned to the cooling cycle via a repumper at 935~nm.  Once every few minutes, the ion can also decay to the $4f^{13}6s^{2}\, {}^{2}F_{7/2}$ state, and must be repumped at 638~nm. The cooling laser and both repumpers have isotope shifts of a few GHz \cite{Martensson-Pendrill1994,Feldker2018,Sugiyama2000}.
	
\begin{figure}
		\centering
		\includegraphics[width=0.9\columnwidth]{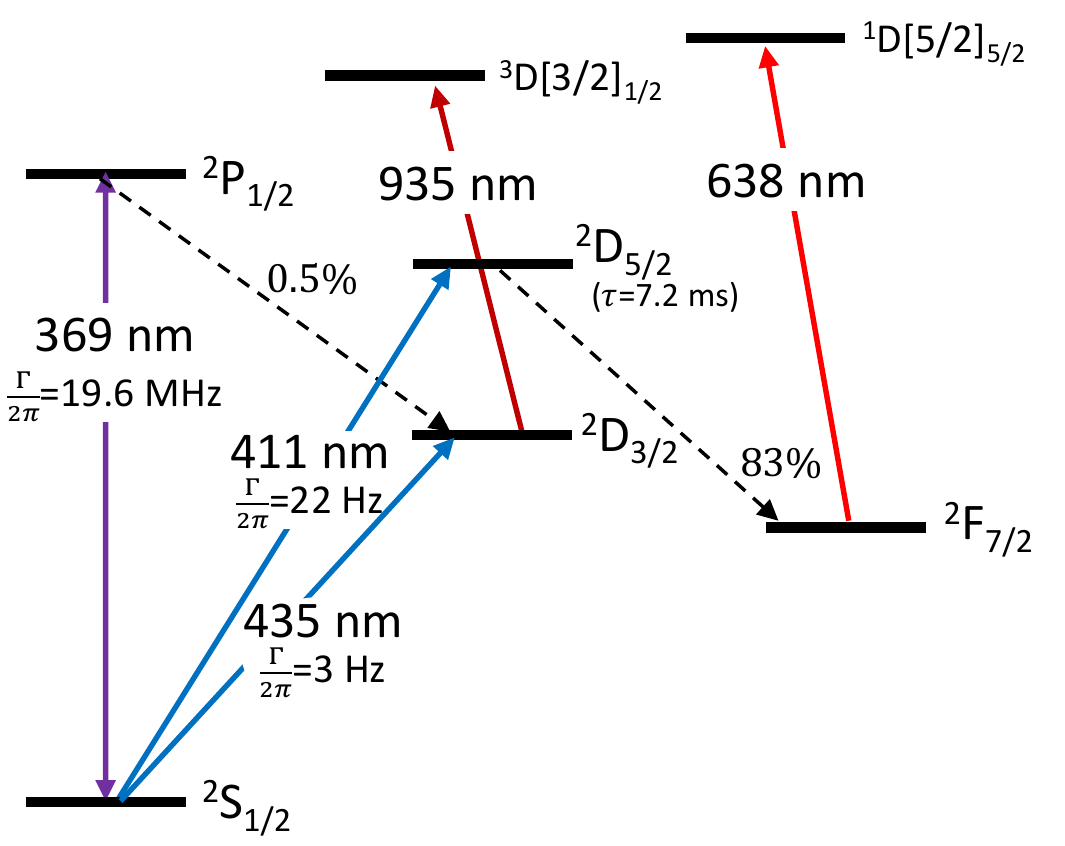}
		\caption{\label{fig:levels}Partial \Yb\, level diagram.}
	\end{figure}

	To drive the probe transitions at 411~nm ($6s\,{}^{2}S_{1/2} \rightarrow 5d\,{}^{2}D_{5/2}$; $\Gamma/(2\pi)=22$ Hz \cite{Taylor1997}) and 436~nm ($6s\,{}^{2}S_{1/2} \rightarrow 5d\,{}^{2}D_{3/2}$; $\Gamma/(2\pi)=3$ Hz \cite{Gerz1988}), we employ a Ti:Sapphire probe laser, tuned to 822~nm and 871~nm, respectively. This laser is frequency-stabilized via the Pound-Drever-Hall (PDH) protocol to an ultra-low-expansion-spaced (ULE-spaced) cavity that has finesse $\mathcal{F} \sim$ 50000 and linewidth $\kappa/(2\pi)=30$~kHz. By frequency-stabilizing a sideband produced by an electro-optic modulator (EOM) to the cavity, a coarse spectroscopic frequency scan can be engineered via tuning the sideband frequency. The infrared light is fiber-coupled to one of two potassium titanyl phosphate (KTP) waveguide doublers, which output 411~nm and 436~nm light, respectively. The blue light, of linewidth $\sim$1 kHz, is then passed through an acousto-optic modulator (AOM), which can be used for finer frequency tuning, and focused down through an achromatic lens to a beam waist of radius $w_0 = $ 15 $\mu$m at the trapped ion. In order to align the probe beam to the ion, 369-nm light is overlapped with the probe.  The 369-nm and the probe beams are focused through the same achromatic lens, and alignment is verified by the resulting fluorescence re-radiated off the ion.
	The powers of the 411 and 436~nm beams at the location of the ion are 1.1~mW and 0.2~mW, respectively. The relevant levels, lasers (cooling, probe, and repumper lasers), and decays are illustrated in Fig.~\ref{fig:levels}.
	
	The readout of the state is carried out via an \textit{electron-shelving scheme} \cite{Taylor1997}. In order to probe the ion in the dark and avoid AC Stark shifts, the 369~nm cooling light and the probe light (411~nm or 436~nm) are alternatively applied. Ions in the ground state $S_{1/2}$ are detected via fluorescence during the cooling with 369~nm light. If the ion is fluorescing before a probe pulse and no longer fluorescing afterwards, the ion is said to have completed a quantum jump. Otherwise, the ion failed to quantum jump (or, if there was no fluorescence before the probe pulse, the ion failed to be initialized). By dividing the number of quantum jumps by the total number of successful initialization, we can measure a probability of excitation as a function of frequency. The details of the pulse sequences used to implement this protocol for the 411~nm and the 436~nm transitions are described in sections~\ref{section:411} and \ref{section:436}, respectively.
	
\begin{figure}
		\centering
		\includegraphics[width=\columnwidth]{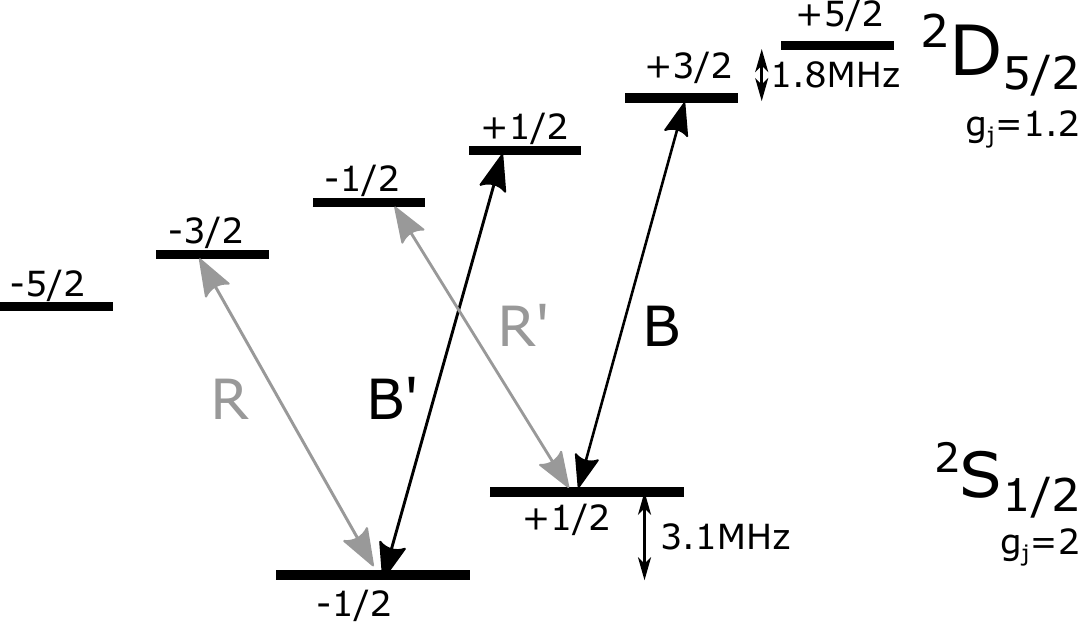}
		\caption{The Zeeman transition pair measured to determine the 411~nm transition center is labelled $R$ and $B$. Estimated Stark shifts due to off-resonant driving of transitions $R'$ and $B'$ are listed in table~\ref{tab:laser-induced_light_shifts}.}
		\label{fig:411_zeemans}
		% \end{figure}
		\vspace{0.6cm}
		% \begin{figure}
		\centering
		\includegraphics[width=0.7\columnwidth]{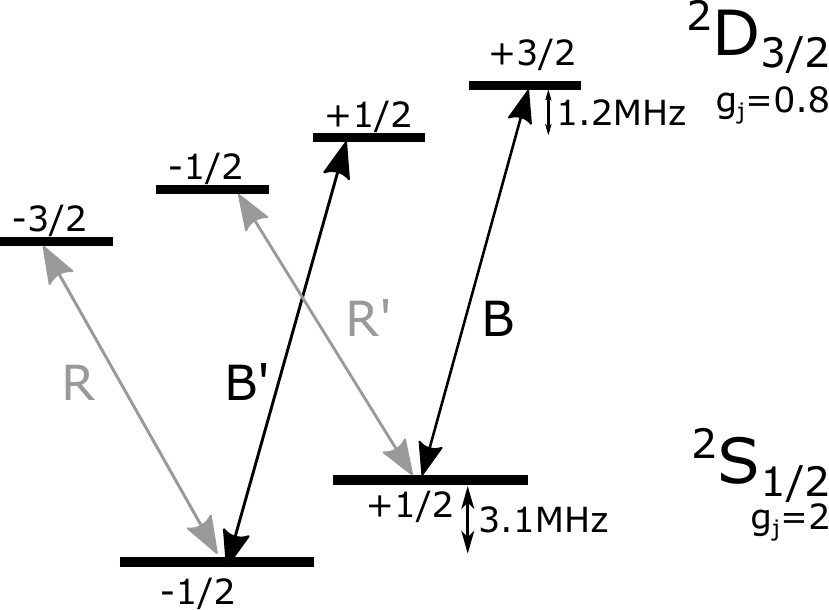}
		\caption{The Zeeman transition pair measured to determine the 436~nm transition center is labelled $R$ and $B$. Estimated Stark shifts due to off-resonant driving of transitions $R'$ and $B'$ are listed in table~\ref{tab:laser-induced_light_shifts}.} 
		\label{fig:436_zeemans}
	\end{figure}

\begin{figure}
		\centering
		\includegraphics[width=\columnwidth]{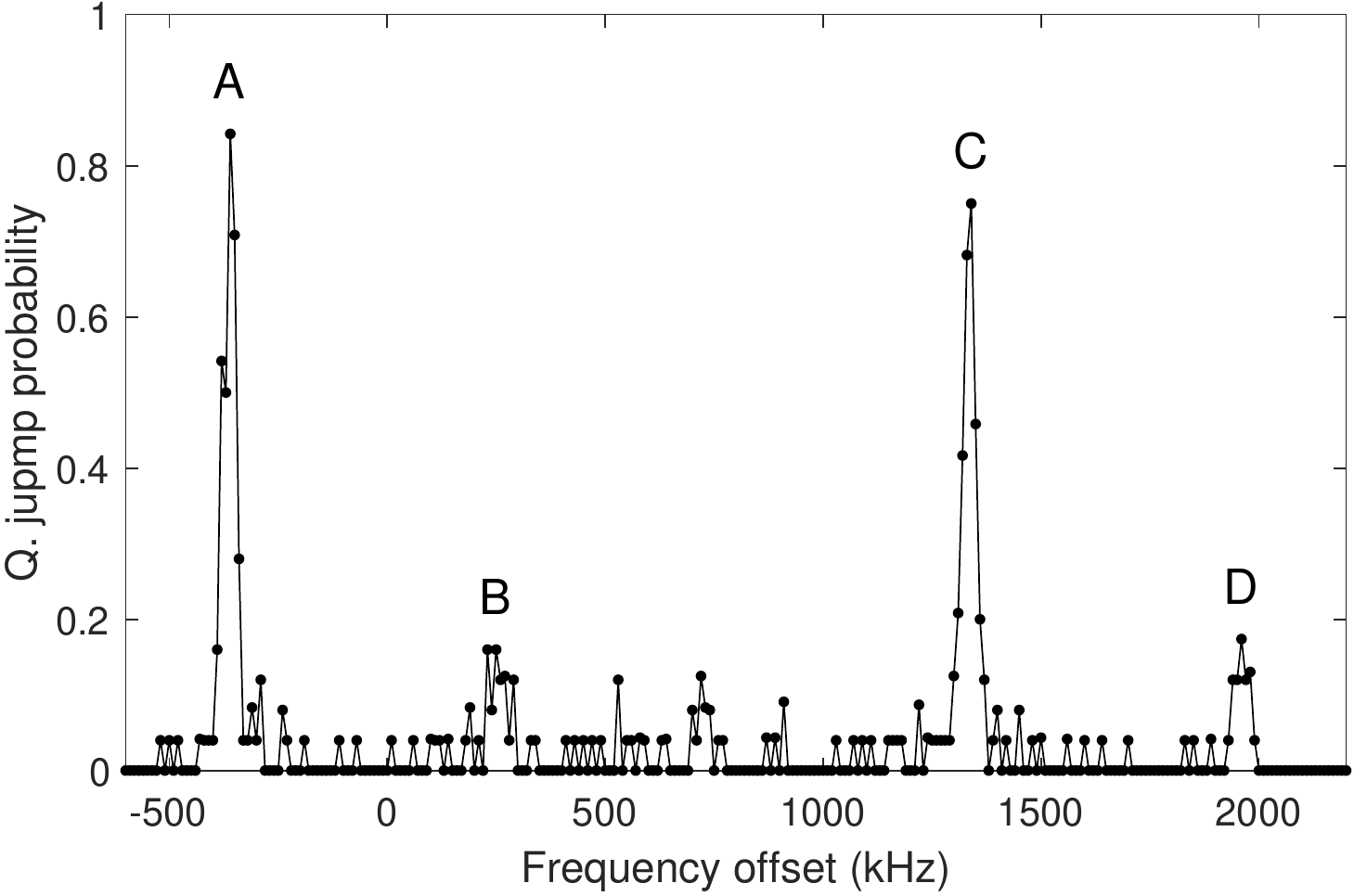}
		\caption{\label{fig:widescan}A broad spectrum taken across four Zeeman components of the 411~nm transition, probed via a single pulse. The Zeeman shift agrees with calculated g-factors for the 411~nm transition and the magnetic field magnitude of 1.1~G. Note that optical pumping suppresses two of the Zeeman components and enhances their respective opposites.  The components shown are: \\
			(A) $\ket{S_{1/2},m_J=1/2} \rightarrow \ket{D_{5/2},m_J=-1/2}$,\\
			(B) $\ket{S_{1/2},m_J=-1/2} \rightarrow \ket{D_{5/2},m_J=-3/2}$,\\
			(C) $\ket{S_{1/2},m_J=1/2} \rightarrow \ket{D_{5/2},m_J=3/2}$,\\
			(D) $\ket{S_{1/2},m_J=-1/2} \rightarrow \ket{D_{5/2},m_J=1/2}$.}
		% \end{figure}
		\vspace{0.35cm}
		% \begin{figure}
		\centering
		\includegraphics[width=\columnwidth]{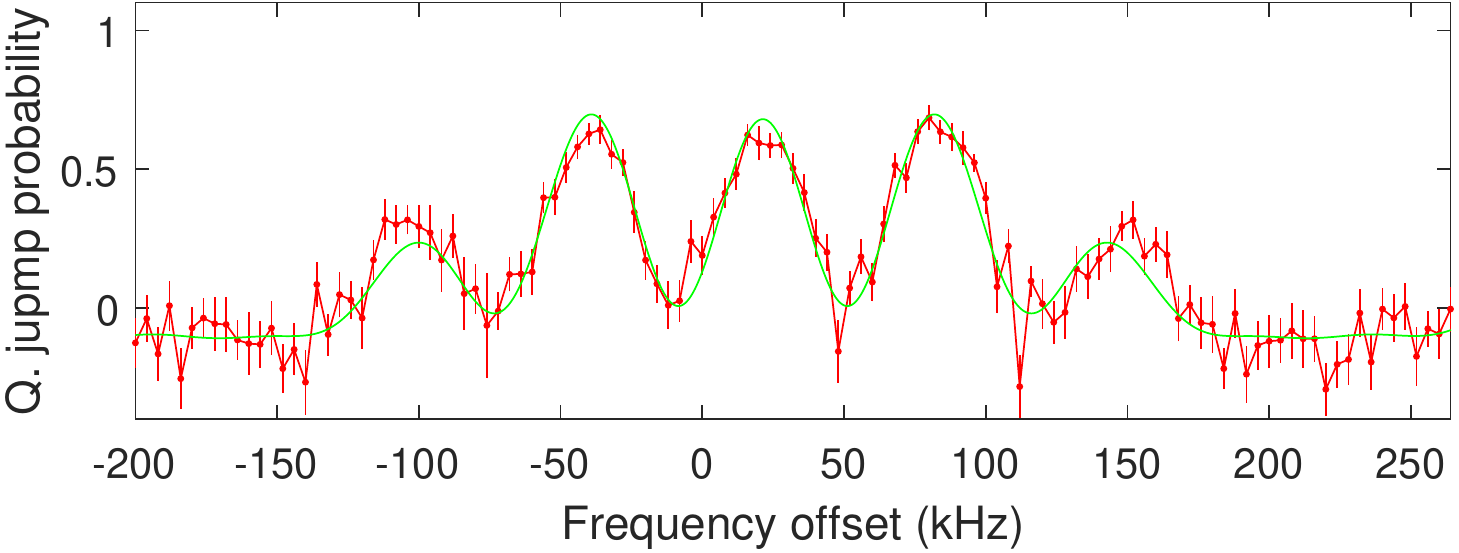}
		\caption{\label{fig:Ramsey}Ramsey spectrum on a single Zeeman component of the 436-nm transition with fit (green solid curve). The x-axis is the frequency of the probe beam with an arbitrary offset. The quantum jump probability is obtained by taking the fractional difference between the integrated readout fluorescence and the integrated calibration fluorescence; see section~\ref{section:436}.}
		% \end{figure}
		\vspace{0.35cm}
		% \begin{figure}
		\centering
		\includegraphics[width=\columnwidth]{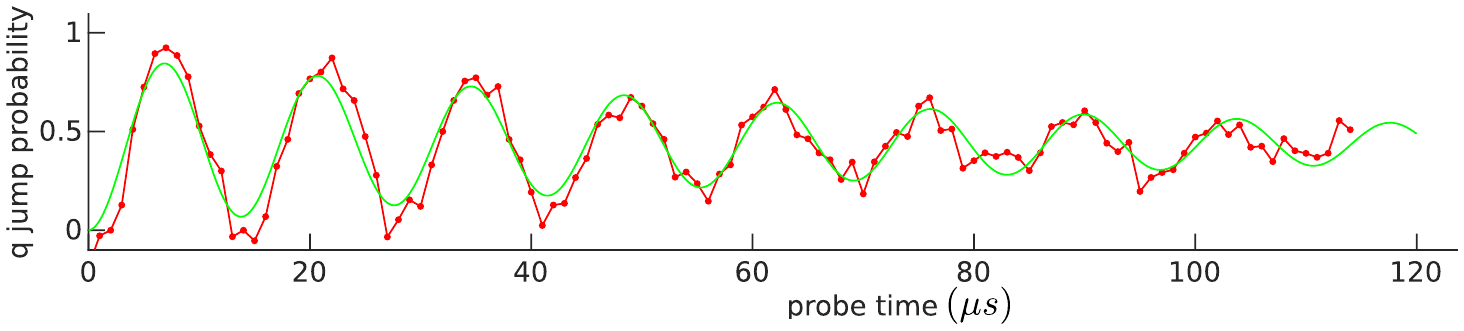}
		\caption{\label{fig:Rabi} Rabi oscillations for a Zeeman component of the 436-nm transition. The measured quantum jump probability is plotted against the scanned probe pulse length. The Rabi oscillation is fitted to determine the $\frac{\pi}{2}$ time, which is used in the Ramsey sequence that we apply to probe the transition. The quantum jump probability is obtained by taking the fractional difference between the integrated readout fluorescence and the integrated calibration fluorescence; see section~\ref{section:436}.}
	\end{figure}

	The precise frequency of either clock transition is determined via Ramsey spectroscopy, with $\frac{\pi}{2}$ times of $\sim5$~$\mu$s and interrogation times of 10~$\mu$s. A small magnetic field is applied to the ion to compensate for Earth's magnetic field, and an additional magnetic field of $\sim1.1$~G is applied along the probe beam to separate different Zeeman components of the transition. The probe laser frequency is scanned (in steps of $2-4$~kHz) over the central Ramsey fringe of a pair of transitions, symmetrically detuned from the center frequency, between Zeeman components of the ground and excited states. This pair of transitions, labelled $B$ and $R$ as indicated in Figs.~\ref{fig:411_zeemans} and Fig.~\ref{fig:436_zeemans}, is identified via a wide scan with a single pulse of the probe beam applied in each measurement cycle (see Fig.~\ref{fig:widescan}, for instance).  Fig.~\ref{fig:Ramsey} shows an example Ramsey spectrum of one Zeeman component of the 436~nm transition. Fig.~\ref{fig:Rabi} shows a probe-pulse-length scan performed on the transition (Rabi flopping), which is used to determine the $\pi/2$ time to be used in the Ramsey sequence.
	
	Five scans over each of the red-detuned and blue-detuned symmetric Zeeman transitions (transitions $R$ and $B$ respectively in Figs.~\ref{fig:411_zeemans}, \ref{fig:436_zeemans}) are interleaved. Each set of scans is repeated two to three times for each isotope. We establish a center frequency for the probed transition by averaging the frequencies of $R$ and $B$. For each measurement of $B$'s  frequency, a pair of measurements of $R$'s frequency taken before and after the measurement of $B$ are used to interpolate the frequency of $R$ at the time when $B$ was measured. The same is done for each measurement of $R$ (using pairs of measurements of $B$ taken before and after the measurement of $R$). Measurements of transition center are then determined by averaging the frequency of the $B$ ($R$) and the interpolated frequency of $R$ ($B$).
	
	To measure the isotope shift between two isotopes $i$ and $j$, a single ion of isotope $i$ is first selected and loaded by tuning the frequency of a photo-ionizing beam aligned to a stream of neutral Yb atoms emitted from an oven. Cooling and repumping beams are also tuned as required. Once isotope $i$ is loaded, the Ramsey measurement described in the previous paragraph is repeated multiple times. The lasers are then re-tuned to load isotope $j$. The probe beam is tuned (but is kept locked to the same cavity free spectral range (FSR)), and the Ramsey measurements are again carried out. The measurements on $i$ and $j$ are then repeated, with each isotope measured between two and three times.  While the absolute frequencies of the clock transitions for different isotopes are not measured independently, this method allows the measurement of a relative isotope shift with precision on the order of $\sim$300~Hz (see Fig.~2 and Table I in the main text). The isotope shifts of four nearest pairs of stable even Yb isotopes (i.e., $j=168,170,172,174$ and $i=j+2$) were measured for each transition. By comparing this precision shift measurement with a previously measured absolute frequency for the $^{172}$Yb$^+$ isotope \cite{Rana2012}, absolute frequencies for the 411~nm transition can also be determined (Table II in the main text).
	
	%Give some details of trap parameters, RF frequency, vibration frequencies, micromotion compensation (?) etc.
	
	\subsection{Measurements on the ${}^2S_{1/2} \rightarrow {}^2D_{5/2}$ transition at $\alpha=411$~nm}
	\label{section:411}
	%To determine the center frequency of the 411 transition, we probe a pair  of transitions, symmetrically detuned from the center frequency, between Zeeman components of the ground and excited states. Here, we use the $S_{1/2}\ket{m_j =-\frac{1}{2}} \rightarrow D_{5/2}\ket{m_j=\frac{x}{x}}$ and $S_{1/2}\ket{m_j =\frac{1}{2}} \rightarrow D_{5/2}\ket{m_j=\frac{x}{x}}$ transitions, which we will refer to as the red-411 and blue-411.
	
	%Three frequency scans of the probe laser are performed over the red-411 and blue-411 transitions are performed in seq $\sigma^+$  polarized 370-nm beam is used to initialize the ion in either the $m_j = +\frac{1}{2}$ level of the $S_{1/2}$ and the probe frequency is scanned over ($sigma^-$)$m_j = -\frac{1}{2}$ 
	
\begin{figure}[h!]
		\centering
		\includegraphics[width=\columnwidth]{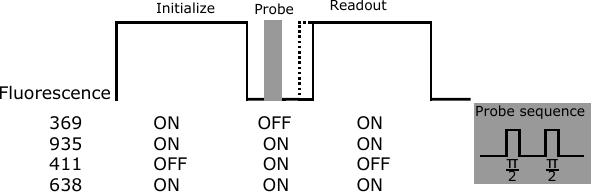}
		\caption{Pulse sequence used to probe the 411-nm transition.}
		\label{fig:seqSingleShot}
	\end{figure}

	The pulse sequence applied at each point of the frequency scans used to probe the 411~nm transition is depicted schematically in Fig.~\ref{fig:seqSingleShot}. The sequence is comprised of three sections: initialization, probe and readout. Its total duration is 200~ms. During initialization, the ion is Doppler-cooled on the 369~nm transition. After cooling, a 40~$\mu$s optical pumping pulse is used to initialize the ion in the $m_j = -\frac{1}{2}$ or $m_j = +\frac{1}{2}$ level of the $^2S_{1/2}$ ground state with $\sigma^-$ or $\sigma^+$-polarized 369~nm light, respectively. This is followed by the probe period, where a Ramsey sequence is applied using 411~nm light (the Ramsey interval used was 10~$\mu$s, and the $\frac{\pi}{2}$-pulse length was $\sim5~\mu$s). During the Ramsey sequence, the 369~nm is extinguished by an AOM. The probe sequence is followed by a readout pulse of 369~nm light, during which the ion's time-resolved florescence is measured and recorded by a photomultiplier tube (PMT) synchronized to a field-programmable gate array-based (FPGA-based) data acquisition system. The sum of the fluorescence counts recorded over the first three 2~ms bins in the readout part of the sequence is then compared to the sum of the fluorescence counts recorded over the last three 2~ms bins in the initialization part of the sequence. The fluorescence counts are then compared to a previously-established threshold. If above the threshold, the ion is taken to be in the ground state and is labelled as ``bright". If the initialization counts do not indicate a ``bright" ion, this implies initialization has failed and there was no opportunity for a quantum jump to occur, since the ion was not in the ground state before the probe sequence. In this case, the sequence is not counted towards the total number of shots.  A quantum jump is taken to have occurred if the ion is dark at readout (i.e. readout counts are equal to or below threshold) and bright at initialization. The sequence is repeated 35 times and a quantum-jump probability is established by taking the ratio of the number of quantum jumps that occurred to the number of attempts with successful initialization. 
	
	It should also be noted that the 638-nm repumper light is kept on during the initialization stage in order to repump any population that may have decayed from $^2D_{5/2}$ to $^2F_{7/2}$ (branching ratio = 83\% \cite{Taylor1997}). The main cause of initialization failure is unsuccessful repumping of this population to the $^2S_{1/2}$ ground state by the 638-nm repumper, which has a repumping time constant of $\gtrsim200$~ms.

	\subsection{Measurements on the ${}^2S_{1/2} \rightarrow {}^2D_{3/2}$ transition at $\beta=436$~nm}
	\label{section:436}
	The pulse sequence with single-shot readout of quantum jumps that is used for probing the 411-nm transition cannot be used when probing the 436-nm transition. This is because the upper state of the 436-nm transition can be populated not only by a probe-induced quantum jump from the $^2S_{1/2}$ ground state, but also by spontaneous decay from the $^2P_{1/2}$ excited state. Single-shot readout of the fluorescence on the 369~nm transition would not distinguish between these two scenarios. 
	
\begin{figure}[h!]
		\centering
		\includegraphics[width=\columnwidth]{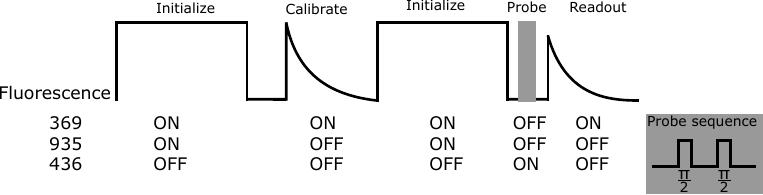}
		\caption{Pulse sequence used to probe 436~nm transition. The solid curve indicates the time-resolved fluorescence by the ion.}
		\label{fig:seqAveraging}
	\end{figure}

	Instead, we employ the pulse sequence depicted in Fig.~\ref{fig:seqAveraging} to probe this transition. The first half of the sequence is comprised of an initialization pulse with both the 369-nm and the 935-nm lasers applied, as before, followed by a time-resolved fluorescence readout with only the 369-nm beam. This readout serves to calibrate the effect of population decay from $^2P_{1/2}$ to the ${}^2D_{3/2}$ -- without the 935-nm repumper, population will eventually be pumped to the ${}^2D_{3/2}$ dark state. The same initialization and readout sequence is then repeated after the Ramsey probe sequence has been applied. When averaged over several applications of the sequence (we perform 2000 shots of the sequence, which corresponds to a 1s integration time for a 500-$\mu$s-long sequence; 8-10 repeats of this 1~s integration time are then performed), this second readout will show a smaller peak height in fluorescence if the 436~nm transition is being driven. A quantum jump probability is hence obtained by taking the fractional difference between the averaged fluorescence counts integrated over all time bins of the second readout and the counts integrated over all time bins of the first calibration readout.
	
	\section{Data Analysis}\label{section:analysis}
	A set of measured center frequencies of a transition for a pair of isotopes over the course of $\sim10$~hours are fitted with the linear model: 
	\begin{equation}\label{eq:ISfit}
	y_k = a + b t_k + c z_k
	\end{equation}
	where $k$ indexes each data point; $y_k$ is the measured transition frequency; $t_k$ is the time at which point $k$ was measured; $z_k = 1$ if the point $k$ is for isotope $j$ and $z_k = 0$ for reference isotope $i$; and $a$, $b$, and $c$ are fitting parameters. The first two terms account for the linear drift of the length of the reference cavity. The last term describes the isotope shift; with $z_k$ set to 0 or 1, $c$ represents the fitted isotope shift.
	
\begin{figure}
		\centering
		\includegraphics[width=\columnwidth]{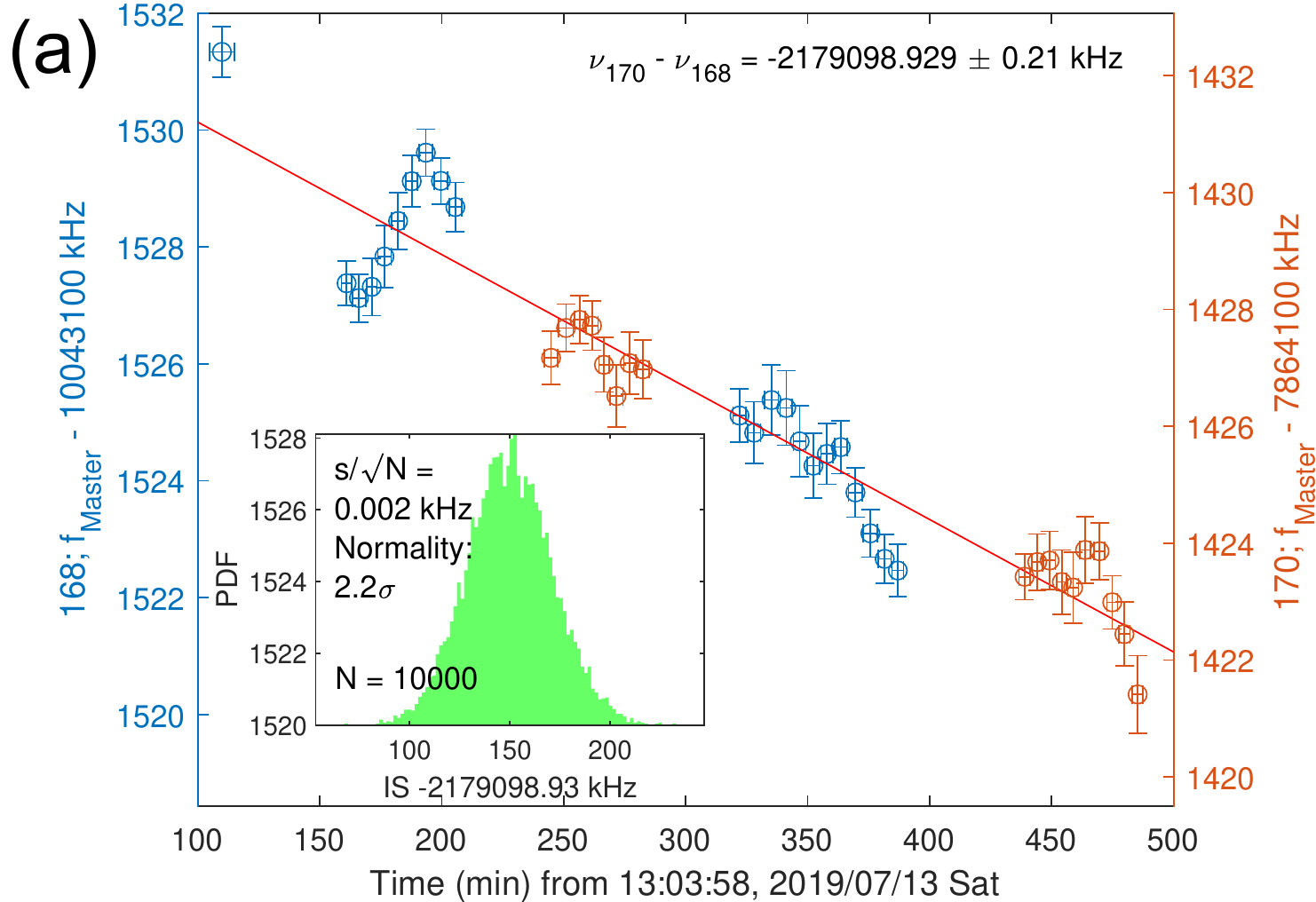}
		\includegraphics[width=0.94\columnwidth]{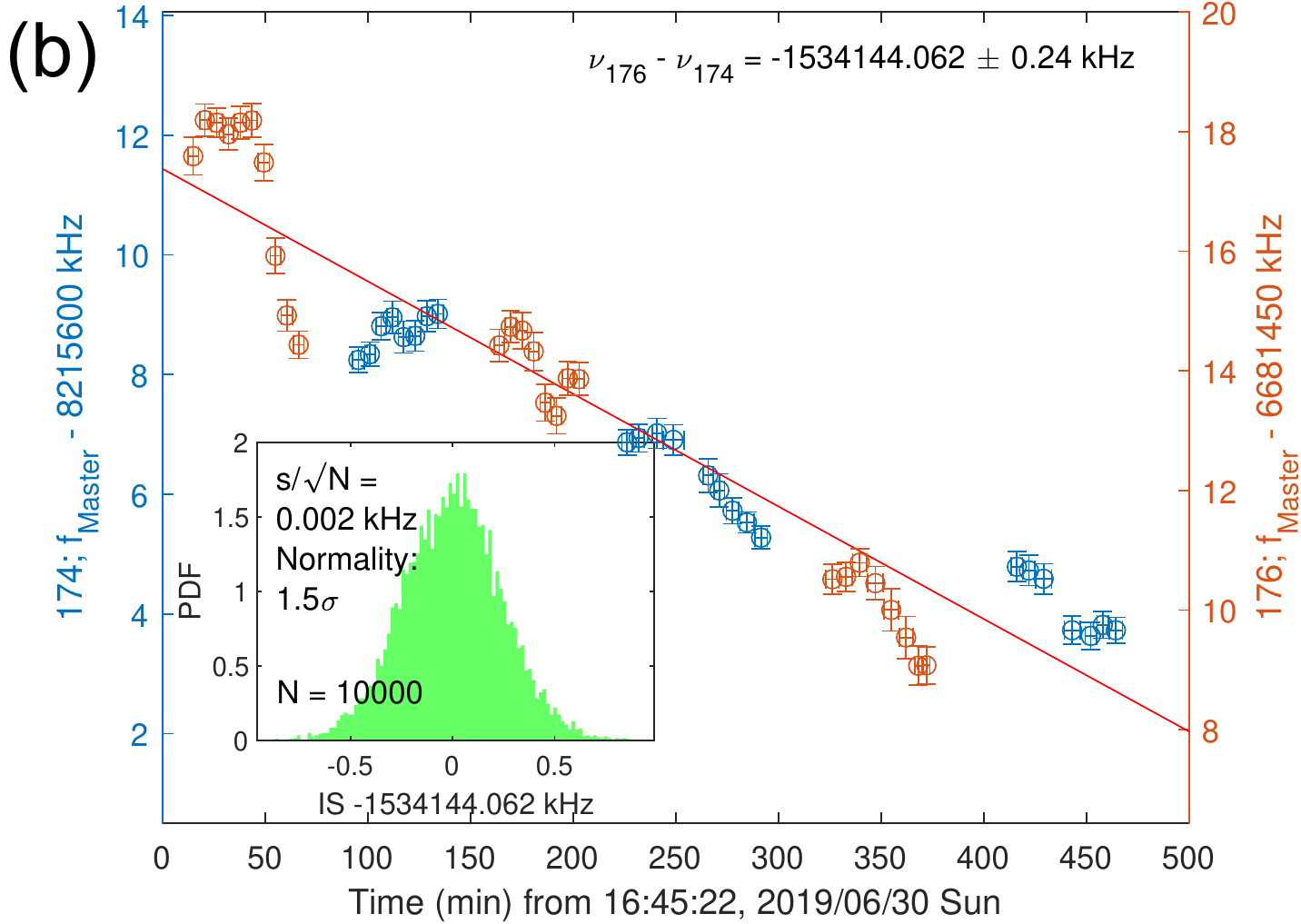}
		\caption{Frequencies of (a) 411~nm transition in $^{168}$Yb$^+$ (blue; left y-axis) and $^{170}$Yb$^+$ (orange; right y-axis), and (b) 436~nm transition in $^{174}$Yb$^+$ (blue; left y-axis) and $^{176}$Yb$^+$ (orange; right y-axis). %An FSR of the reference cavity was chosen for a pair of isotopes and the center frequencies were measured from the FSR.
			Transition frequencies of both isotopes over time were fitted with linear model (red line, see Eq.~\ref{eq:ISfit}). The insets show the distribution of fitted isotope shifts from bootstrapping with $N=10^4$ re-sampling.}
		%Measured isotope shift and its error from mean and standard deviation of the distribution, respectively, are shown below the top x-axis in each graph.Red solid line in each graph shows the fitted drift of the measured transition frequency from the reference FSR. See section \ref{section:analysis} for details.
		\label{fig:ISfit}
	\end{figure}

	To determine the uncertainty in the measured isotope shifts, bootstrapping statistics are employed \cite{[{Chapter 9: Regression models in }] [{. {\textit{Bootstrapping pairs}} was used, see Chapter 9.5 in the book.}] Efron1993}. A new set of data points is formed by re-sampling points from the set of measured data points, allowing for multiple instances of the each point, until the number of elements in the re-sampled set is the same as that of the original set. The fitting described in the previous paragraph is applied to the re-sampled set to find a value for the isotope shift. Repeating the procedure $N$ times (with sufficiently large $N$) gives the histogram of isotope shifts from each re-sampled set. The mean and standard deviation of the distribution are taken as the measured isotope shift and its uncertainty, respectively. Fig.~\ref{fig:ISfit} shows examples of the measured transition frequencies over time, the fitted linear drift of the reference FSR, and the bootstrapping statistics (see insets). Drifts in the probe laser frequency with time are visible in Fig.~\ref{fig:ISfit}. We discuss estimates of systematic shifts in section \ref{section:systematics}.
	
	The isotope shifts between nearest even isotopes are measured for each transition. Additionally, the isotope shift between ${}^{170}$Yb and ${}^{174}$Yb is measured as a cross-check. The results are listed in table~I in the main text. Using the additional $(j,i) = (170,174)$ measurement, the precision of the isotope shifts $\nu_{\alpha ji}$'s for $(170,172)$, $(172,174)$, and $(170,174)$ can be improved in the standard way by adding and averaging measured values:
	\begin{equation}
	\begin{aligned}
	\nu'_{1} &= \frac{\sigma'^{2}_1}{\sigma^2_1}\nu_1 + \frac{\sigma'^{2}_1}{\sigma^2_2 + \sigma^2_3}(\nu_3-\nu_2) \quad &\frac{1}{\sigma'^{2}_1} = \frac{1}{\sigma^2_1} + \frac{1}{\sigma^2_2 + \sigma^2_3} \\
	\nu'_{2} &= \frac{\sigma'^{2}_2}{\sigma^2_2}\nu_2 + \frac{\sigma'^{2}_2}{\sigma^2_3 + \sigma^2_1}(\nu_3-\nu_1) \quad &\frac{1}{\sigma'^{2}_2} = \frac{1}{\sigma^2_2} + \frac{1}{\sigma^2_3 + \sigma^2_1} \\
	\nu'_{3} &= \frac{\sigma'^{2}_3}{\sigma^2_3}\nu_3 + \frac{\sigma'^{2}_3}{\sigma^2_1 + \sigma^2_2}(\nu_1+\nu_2) \quad &\frac{1}{\sigma'^{2}_3} = \frac{1}{\sigma^2_3} + \frac{1}{\sigma^2_1 + \sigma^2_2}, \\
	\end{aligned}
	\end{equation}
	where $\nu_q$ and $\sigma_q$ refer to the measured shift $\nu_{\alpha ji}$ and its uncertainty, respectively, for $q=1$: $(170,172)$, $q=2$: $(172,174)$, and $q=3$: $(170,174)$ pairs, and $\nu'_q$ and $\sigma'_q$ represent the values with the $(170,174)$ measurement included. The results are shown in table~\ref{table:improved_IS}. Through this procedure, however, the isotope shifts $\nu'_1$, $\nu'_2$, and $\nu'_3$ become correlated to each other. To take the correlation into account, the generalized-least-squares (GLS) method is used to fit the points in King plot (Fig. 2 in the main text).
	We extract the $p$-value of the $\chi_{k}^2$ distribution, with $k=2$ being the degree of freedom of the fitting (for four data points and two fitting parameters). We find $\chi^2 = 11.7$ with a $p$-value of 0.0029, corresponding to a significance of 3 $\sigma$.
	
	There are different ways of choosing the reference isotope, and the choice determines the correlation between errors. When the latter are included carefully in the analysis, the same significance of nonlinearity, within $\pm 0.001~\sigma$, is obtained for all reference choices. The confidence intervals of the nonlinearity measures, parametrized by $\upsilon_{ne} D_{\beta\alpha}$ and $G_{\beta\alpha}$ (Fig.~3b in the main text), from different reference choices also agree.
	
	To obtain the nonlinearity pattern from the known possible sources (e.g., $G_\alpha \drtsq_{ji}$ or $\upsilon_{ne} D_\alpha a_{ji}$), the four points $(\overline{\nu}_{\alpha ji},z_{ji})$ are fitted in the same way the King plot is fitted, where $z_{ji}$ is the nuclear factor of the source term (e.g. $\drtsq_{ji}$ or $a_{ji}$). The residuals of the fitting show the nonlinearity pattern of the source, and the corresponding nonlinearity measure $\zeta_\pm^z$ is obtained via Eq.~4. The direction of the nonlinearity lines is given as the ratio $\zeta_-^z/\zeta_+^z$, as in Fig.~3a. Note that the $pattern$ is determined by nuclear factors while the electronic factors only scale the overall size of residuals.

	\begin{table}
		\caption{\label{table:improved_IS}Isotope shifts $\nu'_{ji}$'s between $(j,i=j+2) = (170,172)$, (172,174), and (170,174) isotope pairs after the precision is improved. The values are correlated to each other for each transition.}
		\begin{ruledtabular}
			\begin{tabular}{ccc}
				Isotope pair & $\nu'_{\alpha ji}$ [kHz] & $\nu'_{\beta ji}$ [kHz] \\
				$(j, i)$ & $\alpha: {}^2S_{1/2} \rightarrow {}^2D_{5/2}$ & $\beta: {}^2S_{1/2} \rightarrow {}^2D_{3/2}$ \\
				\hline
				(170, 172) & 2 044 854.73(30) & 2 076 421.04(28) \\
				(172, 174) & 1 583 068.35(31) & 1 609 181.29(20) \\
				(170, 174) & 3 627 923.08(35) & 3 685 602.34(27)
			\end{tabular}
		\end{ruledtabular}
	\end{table}

	\section{Suppressed effect of mass uncertainty}
	
	The uncertainties in $\overline{\nu}_{\alpha ji}$ and $\overline{\nu}_{\beta ji}$ due to the uncertainties in $\mu_{ji}$ from the measured masses are correlated as the change in $\mu_{ji}$ modifies both $\overline{\nu}_{\alpha ji}$ and $\overline{\nu}_{\beta ji}$. If the ratio of the change
	\begin{equation}
	\frac{\partial\overline{\nu}_{\beta ji}/\partial\mu_{ji}}{\partial\overline{\nu}_{\alpha ji}/\partial\mu_{ji}} = \frac{\nu_{\beta ji}}{\nu_{\alpha ji}}
	\end{equation}
	is parallel to the slope of the King plot $F_{\beta \alpha}$, the effect of mass uncertainty will be suppressed. The suppression factor is given as the difference between the ratio $\nu_{\beta ji}/\nu_{\alpha ji}$ and the slope $F_{\beta \alpha}$
	\begin{equation}
	\begin{aligned}
	\frac{\nu_{\beta ji}}{\nu_{\alpha ji}} - F_{\beta \alpha} &= \frac{F_\beta \drt_{ji} + K_\beta \mu_{ji}}{F_\alpha \drt_{ji} + K_\alpha \mu_{ji}} - \frac{F_\beta}{F_\alpha} \\
	&= F_{\beta \alpha} \left[\frac{K_\beta \mu_{ji}}{F_\beta \drt_{ji}} - \frac{K_\alpha \mu_{ji}}{F_\alpha \drt_{ji}}\right] \\
	&\quad + \mathcal{O}\left[\left(\frac{K_\alpha \mu_{ji}}{F_\alpha \drt_{ji}}\right)^2\right]
	\end{aligned}
	\end{equation}
	It is known that mass shifts are smaller than field shifts for heavy elements in general. For the $\alpha=411$~nm and $\beta=436$~nm transitions in Yb$^+$, the ratio of mass shift and field shift (FS) is $\sim 5$\% (see table~\ref{table:ChargeRadius}). There is even further suppression due to the similar ratio for the transitions $\alpha$ and $\beta$, giving a suppression factor $\sim 0.1\%$. The uncertainty in the measured mass of ${}^{168}$Yb, 1.3~$\mu$u, corresponds to the uncertainty $\mathcal{O}\mathrm{(1~kHz)}$ in $\nu_\alpha$ or $\nu_\beta$, and $\mathcal{O}\mathrm{(1~Hz)}$ after the effect on the linear fit is suppressed. The mass uncertainties of the other isotopes $\mathcal{O}\mathrm{(0.01~\mu u)}$ have an effect $\mathcal{O}\mathrm{(10~mHz)}$ on the linear fit. When this level of frequency precision is reached in future precision spectroscopy, it will be necessary to improve the mass measurements.
	
	\section{Estimation of systematic effects and errors}\label{section:systematics}
	The uncertainties in our measurement are determined directly from the variation of our data points, as described in section \ref{section:analysis}. In the following subsections, we estimate the magnitude of the physical effects we expect to have contributed to this uncertainty, and of any additional systematic shifts. Table \ref{tab:systematic_shifts} summarizes these estimates. Most of the effects discussed here are largely common-mode for the isotope pairs we measure, and produce either zero or very small differential systematic shifts in our measurement (we estimate a total systematic differential shift of $<10$\,Hz - see column 3 in table~\ref{tab:systematic_shifts} - which corresponds to $<10\%$ of our statistical uncertainty). However, we discuss here how drifts over time in experimental parameters lead to uncertainties in these differential shifts (listed in column 3 of table~\ref{tab:systematic_shifts}); these uncertainties account for most of our measurement error and are seen in the scatter of the data points. For reference, we also list in column 2 of table~\ref{tab:systematic_shifts}, for each systematic effect, an estimate for the absolute value of the shift it produces on the transition frequency of a single isotope.
	
	\subsection{Second-order Doppler Shift}
	The fractional Doppler shift to the atomic transition, $\Delta\nu_D / \nu_0$, caused by the motion of the ion in the trap is given by:
	\begin{equation}
	\label{eqDoppler}
	\frac{\Delta\nu_D}{\nu_0}= -\cos(\theta)\frac{v}{c}-\frac{v^2}{2c^2}+\mathcal{O}\left({\left(\frac{v}{c}\right)}^3\right)
	\end{equation}
	where $v$ is the absolute instantaneous velocity of the ion relative to the lab frame, $\theta$ is the angle of observation and $\nu_0$ is the frequency of the atomic transition in the rest frame of the ion.
	
	Since we are in the Lamb-Dicke regime, we can ignore the first term of this equation, the linear Doppler shift, $\frac{v}{c}$, because this term will simply add sidebands to the transition but will not shift the carrier \cite{Lizuain2007}.
	
	The second term shifts the atomic transition frequency due to relativistic time dilation. We calculate this term for both micromotion and secular motion of the ion in the trap. Because we are sampling the instantaneous velocity over a time much larger than one oscillation period of the secular or micromotion, we can replace instantaneous velocity by mean square velocity. So the overall fractional second-order Doppler shift is
	\begin{equation}
	\frac{\Delta\nu_D}{\nu_0}= -\frac{\langle v_s^2\rangle}{2c^2}+{\left(\frac{\Delta\nu}{\nu_0}\right)}^{\mu\text{motion}}
	\end{equation}
	where $\nu_0$ is the unshifted transition frequency, $\langle v_s^2\rangle$ is the mean-square velocity for the secular motion, and ${\left(\frac{\Delta\nu}{\nu_0}\right)}^{\mu\text{motion}}$ is the micromotion-induced fractional Doppler shift. The secular-motion term can be estimated from the ion's temperature, which we take to be of order the Doppler limit on the 369~nm cooling transition ($\approx 500\mu$K), giving $\frac{\langle v_s^2\rangle}{2c^2} \approx 1\times 10^{-20}$. The micromotion term can be calculated for our trap parameters using Eq.~30 in \cite{Berkeland1998}. We conservatively estimate that the ion experiences a DC field in the trap $E$ of order $50\,$V/m$\,$ which gives a micromotion-induced fractional shift of $-8\times10^{-17}$. This dominates over the secular motion shift. 
	From this, we estimate that a small systematic differential second-order Doppler shift of 2\,mHz will arise from the mass difference between isotopes. The main source of uncertainty on this differential shift is expected to be temporal drifts in micromotion compensation. If we assume that $E$ can change by around $50\,$V/m, between measurements of isotope transition frequencies, we arrive at an uncertainty on this differential second-order Doppler shift of $\approx 100$mHz.
	
	\subsection{Black-body Shift}
	The black-body radiation (BBR) shifts on the transitions probed here are well approximated by \cite{Safronova2012}:
	\begin{equation}
	\Delta\nu_\text{BBR} = -\frac{1}{2}\Delta\alpha_0(831.9~\text{V/m})^2{\left(\frac{T}{300~\text{K}}\right)}^4
	\end{equation}
	where $\Delta\alpha_0$ is the difference in scalar polarizability between the atomic states associated with the transition of interest.
	
	Calculations of the fractional BBR shift for the 436~nm transition in $^{171}\text{Yb}$ have estimated $\Delta\nu_\text{BBR}\approx -0.4$~Hz \cite{Safronova2012,Roy2017}. We assume similar results for the 411~nm transition since the similar orbital wavefunctions of the $^2D_{3/2}$ and $^2D_{5/2}$ states imply that $\Delta\alpha_0$ should be similar for both transitions (the small difference in the wavefunctions arises from the relativistic effect). The main source of a differential BBR shift in our experiment will be temperature drifts. Conservatively, we estimate that the temperature can change by up to 3~K over the course of our shift measurement, which yields a change in $\Delta\nu_\text{BBR}$ of $\approx20$~mHz.
	
	\subsection{Electric quadrupole shift}
	A frequency shift results from the interaction of the quadrupole moment of the electronic state with electric field gradients from the trap. The shift is of order
	\begin{equation}
	\label{eq:quadshift}
	\Delta\nu_\text{quad} \sim \frac{\Theta\cdot\nabla E}{h}
	\end{equation}
	The quadrupole moments for the $^2D_{3/2}$ and $^2D_{5/2}$ states of $\Yb$ have been calculated to be $2.068(12)ea_0^2$  and $3.116(15)ea_0^2$ respectively \cite{Nandy2014}. Time-varying electric field gradients due to patch potentials on the chip trap can lead to a differential shift between isotopes. We observe a typical day-to-day variation the DC micromotion compensation voltages applied to our trap electrodes of 20\,mV. Conservatively, we consider a maximum variation of 200\,mV during the course of a shift measurement data-taking run. From this, we infer that differential patch-potential gradients of order $\lesssim1~\text{V/}{\text{mm}}^2$ could occur, which would lead to a differential quadrupole shift of $\lesssim 2$\,Hz.
	
	\subsection{Gravitational red shift}
	Differential gravitational shifts in the measured isotope transition frequencies could arise from changes in height of the apparatus due to vibrations or thermal expansion over the course of our measurements. Considering the thermal expansion of the building our lab is housed in and typical amplitudes of vibrations of optical table surfaces, we estimate that such height changes should be of order 1\,mm, which would lead to a differential shift uncertainty of $\sim0.1$\,mHz.
	
	\subsection{Laser-induced AC Stark shifts}
	\subsubsection{Off-resonant probe light couplings}
	Most AC stark shifts in this experiment are common-mode between isotopes, with differential shifts arising only due to laser-intensity drifts and small fractional frequency differences between the isotopes. The transitions we probe in $\Yb$ share the $^2S_{1/2}$ ground state both with far-off-resonant transitions in the atom (e.g. 369~nm transition $^{2}S_{1/2}\longleftrightarrow~^{2}P_{1/2}$) and the closer-detuned Zeeman components of probed transition (see Figs.~\ref{fig:411_zeemans},~\ref{fig:436_zeemans}). The latter cause significant light shifts, but these are equal and opposite for the each pair of Zeeman-component transitions we measure (i.e. the shift on $R$ is equal and opposite to that on $B$), and will hence largely cancel out after averaging. Calculated estimates of the light shifts caused by off-resonant coupling of the probe laser are listed in table~\ref{tab:laser-induced_light_shifts}. 
	
	We estimate that probe-light intensity fluctuations of order 3\% between measurements of each Zeeman transition will lead to uncertainties on the center of order $30$~Hz. 
	
	We also consider that the probe light intensity can systematically vary by up to 20\% when tuned to different isotope transition frequencies. This can lead to a systematic change in the probe-induced AC stark shift for different isotopes. However, this effect is again largely cancelled out since the shift on the two Zeeman transitions we measure are equal and opposite. The remaining systematic shift will hence arise only from any potential deviation from linear polarization of the probe laser beam, which could cause one of the Zeeman transitions to be preferentially driven. This deviation is limited to 1 part in $10^4$ by a Glan-Taylor polarizer placed in the probe laser beam path in combination with the effect of the vacuum chamber window. Hence, we estimate a systematic differential light shift of $\lesssim 20$\,mHz between isotopes.
	
	\begin{table*}
		\caption{Estimated laser-induced AC Stark shifts due to off-resonant couplings of the probe laser.}
		\begin{ruledtabular}
			\begin{tabular}{m{3cm}  m{7cm}  m{7cm}}
				Off-resonantly driven transition & Estimated Stark shift on 411~nm transition [Hz] & Estimated Stark shift on 436~nm transition [Hz]\\
				\hline
				369& $-1.2\times10^{2}$&$-2.4\times10^2$ \\
				467&$+5.1\times10^{-20}$&$+2.8\times10^{-19}$ \\
				411&---------&$-1.1\times10^{-3}$ \\
				436&$+5.5\times10^{-5}$& --------- \\
				%R&$R'$:$-300$&$R'$:$-330$\\
				%$R$&$B'$:$+300$&$B'$:$+330$\\
				$B'$&$R$:$-800$&$R$:$-1160$\\
				$R'$&$B$:$+800$&$B$:$+1160$\\
			\end{tabular}
		\end{ruledtabular}
		\label{tab:laser-induced_light_shifts}
	\end{table*}
	
	\subsubsection{Non-probe light leakage}
	AOM leakage of 369-nm light during the probe time of the experiment can shift the $^2S_{1/2}$ ground state of the probed transitions. We estimate the leakage to be $\approx 5$~nW (from a $100~\mu$W, 20~MHz red-detuned beam focused to a beam waist of $75~\mu$m), which leads to a shift of $-1.3\times10^3$~Hz. Similarly, we estimate that leakage of the 935~nm beam will shift the excited $^2D_{3/2}$ state of the 436~nm probed transition by $+440$~Hz. Both these shifts are common-mode between isotopes, but intensity drifts of the leaked laser light, estimated to be of order $3\%$, will contribute an uncertainty of order $30$~Hz.
	
	A less significant light shift will also arise from the 935~nm light left on during the 411~nm pulse sequence probe time. The 935~nm beam can also shift the 1070~nm transition connecting the $^2D_{5/2}$ state (excited state of the 411~nm transtion) to the $^3D[3/2]_{1/2}$ state (excited state of the 935~nm transition). We estimate this shift to be of order $-4\times10^{-2}$~Hz, contributing an uncertainty of $\approx1$~mHz to our measurements.
	
	Finally, a 402\,nm laser beam was also used during our experiments to transfer-lock an optical cavity used for increasing the ionization power during ion loading \cite{Cetina2013}. We estimate that the uncertainty from the AC Stark shift caused by this laser is $<10$\,Hz, assuming a maximum intensity drift of $30\%$.
	
	\subsection{Shift of center of Ramsey fringe by off-resonant Zeeman transitions}
	The measured Ramsey fringe of a Zeeman transition of interest is perturbed by other Zeeman transitions that are being driven off-resonantly at the same time. The observed signal can be either a sum of different Ramsey fringes, or there may be quantum interference if the off-resonant transition shares a state with the transition of interest. We can estimate the magnitude of the frequency pulling by fitting a sum of different Ramsey fringes.
	
	The maximum size of the pulling is $\sim 20$~Hz for a detuning $\sim 1$~MHz. The frequency pulling has opposite sign for symmetric Zeeman transitions $R$ and $B$ and the effect will be cancels out after the frequencies of transition $R$ and $B$ are averaged. The differential shift from intensity fluctuation and asymmetric $\sigma^\pm$ polarizations is suppressed due to the fact that Ramsey fringes are insensitive to the change in $\omega_R$ to the first order. The size of the pulling due to $B$-field fluctuation can be significant and $O(10)$~Hz is taken as the upper bound of the effect.
	
	\subsection{Micromotional Stark shift}
	If the ion is shifted off the RF null of the Paul trap by stray DC fields, the RF field it experiences will Stark shift the transitions we probe. This shift is given by \cite{Dube2013}
	\begin{equation}
	\Delta\nu = -\tfrac{\langle E^2\rangle}{2h}\left(\Delta\alpha_0 + \tfrac{1}{2}\alpha_2(3\cos^2\beta-1)\left[\tfrac{3m_j^2-J(J+1)}{J(2J-1)}\right]\right)
	\end{equation}
	where $\langle E^2\rangle$ is the mean-squared value of the electric field experienced by the ion, $\beta$ is the angle between the electric field and the quantization axis, $\Delta\alpha_0$ is the difference in the scalar polarizabilities between the ground and excited states of the transition, $\alpha_2$ is the tensor polarizability of the excited level (the tensor polarizability for the $^2S_{1/2}$ ground state is zero) and $J,m_j$ are the angular momentum quantum numbers for the excited state. 
	
	Based on doubling the typical day-to-day variation we observe in micromotion compensation voltages for our trap, we estimate that the stray DC field experienced by the ion to be $\approx 20$~V/m. This gives a micromotion amplitude of $A_{\mu m}=50$~nm for our trap, which can be translated into an RF electric field amplitude of $E_0 = \frac{m\Omega^2_\text{RF}A_{\mu m}}{e} \approx 24$~V/m (where $\Omega^2_\text{RF}$ is the RF drive frequency in for our trap and $m$ is the mass of the $\Yb$ isotope). Conservatively, we use here $E_0 = 50$\,V/m. From this, and using values for $\Delta\alpha_0$ and $\alpha_2$ for the $^2D_{3/2}$ level from reference \cite{Roy2017}, we estimate a micromotional Stark shift of order 1\,mHz.
	
	The differential shift between isotopes is expected to be dominated by changes in the stray field landscape in the trap during the isotope-shift measurement, and hence we include the full 1mHz in our uncertainty budget for this shift. Note that, even in the absence of stray-field fluctuations, there is a systematic $\approx2\%$ change in the value of this shift between isotopes since the micromotional Stark shift is proportional to the square of the ion mass.
	
	\subsection{AOM switching-induced phase chirp}
	Phase shifts in $\frac{\pi}{2}$-pulses induced when an AOM switches the light are known to cause systematic errors in transition frequencies measured via Ramsey spectroscopy \cite{Degenhardt2005,Sanner2018}. Ref.~\cite{Degenhardt2005} reported the shift in transition frequency by 1.6~Hz when the pulse time $\tau = 1.5~\mu s$ and the interrogation time $T = 21.6~\mu s$ are used for 657-nm transition in Ca. As the pulse time in our experiment is longer (which makes the effect smaller), the interrogation time is of the same order of magnitude, and the frequency of the probe light is similar, the effect is expected to be less than $O(1~\text{Hz})$.

	\subsection{Zeeman shifts and absolute frequency stability of the probe light}
	Both magnetic field drifts and fluctuations in the absolute frequency stability of the probe laser, which we lock to an ultra-low expansion cavity, will lead to frequency shifts of the transitions we probe. Changes in magnetic field that do not occur much faster than the time it takes us to scan over a given Zeeman-component (on the order of a few minutes) should lead to oppositely-signed linear shifts on the measured frequency of Zeeman components symmetrically detuned from the transition center. We focus this discussion on linear Zeeman shifts since, at our magnetic field of 1.1~G, for ions with no hyperfine structure like the even $\Yb$ isotopes we measure here, the quadratic Zeeman shift is expected to be of order $100$~mHz \cite{TrappedChargedParticlesBook}, significantly smaller than the linear shift (based on the measured current noise in our magnetic-field coils, we estimate that our magnetic-field noise is $\lesssim0.1\%$, giving an uncertainty on the quadratic Zeeman shift of 0.2\,mHz.). 
	
	To ascertain whether B-field drifts contributed significantly to the spread in our measured transition centers, we performed measurements of the center using Zeeman components of the excited state with larger B-field sensitivity. We found no significant change in the spread of the data when more B-field sensitive Zeeman states were used. Hence, we conclude that linear Zeeman shifts do not contribute significantly to the point-to-point frequency shifts we observe in our experiment; we estimate here a contribution of the order of $300$~Hz (before averaging over repeated measurements). 
	
	The main contribution to our observed point-to-point frequency drifts derives from fluctuations in the absolute stability of the probe laser locked to the ULE cavity. Residual Amplitude Modulation (RAM) effects from the EOM used for this laser's PDH lock produce a common-mode shift on the measured frequency opposing Zeeman components of the probed transition. As well as RAM, thermal drifts of the cavity due to intracavity light also contribute to fluctuations in the probe laser's absolute frequency stability. We estimate that these effects lead to point-to-point frequency shifts of $\sim1$~kHz in our experiment. This is our largest source of uncertainty, but it is reduced by repeated averaging to a level consistent with the $\sim300$~Hz-spread we observe in our final data that is included in the quoted statistical error.
	
	\begin{table*}
		\caption{Estimated contributions to measurement error. Since we measure differences between transition frequencies of pairs of isotopes, only differential shifts affect our measurement, but absolute shifts are also listed here for reference (column 2). The main contributions to our measurement error come from uncertainties on the differential shifts (column 3), which arise mainly from temporal drifts in experimental parameters between measurements of different isotopes (the systematic differential shifts listed here are estimated for next-neighboring isotopes). The uncertainties listed in column 3 are per measured data point (i.e. if the isotope shift were inferred from a single measurement of the transition center in each isotope). For the dominant shifts, we also provide an estimate of the uncertainty after averaging over ten measurements of the center (for most shift measurements, we perform 15 measurements of the transition center but, for measurements involving the 168 isotope, we perform only 10 repeats due to the long loading times required for this isotope). The estimation of the errors listed in this table is detailed in the text (section~\ref{section:systematics}).}
		\begin{ruledtabular}
			\begin{tabular}{m{5.5cm} m{5cm} m{7cm} }
				&Estimated Magnitude of Absolute Shift [Hz] &Estimated Differential Shift [Hz] \\
				\hline
				Second-order Doppler Shift &$\phantom{0}5\times10^{-2}\pm1\times10^{-1}$&$\phantom{0}2\times10^{-3}\pm1\times10^{-1}$ \\
				Black-body shift &$\phantom{0}4\times10^{-1}\pm2\times10^{-2}$&$\phantom{0\times10^{+0}}0\pm2\times10^{-2}$ \\
				Electric quadrupole shift& $\phantom{0}2\times10^{0\phantom{+}}\pm 2\times10^{0}$&$\phantom{0\times10^{+0}}0\pm 2\times10^{0}$  \\
				Gravitational redshift &$\phantom{0}1\times10^{-2}\pm1\times10^{-4}$&$\phantom{0\times10^{+0}}0\pm1\times10^{-4}$\\
				Laser-induced Stark shift  &$\phantom{0}1\times10^{-1}\pm4\times10^{1}$&$\phantom{0}2\times10^{-2}\pm4\times10^{1}$ \\
				Micromotional Stark shift &$\phantom{0}2\times10^{-1}\pm1\times10^{-3}$&$\phantom{0}2\times10^{-5}\pm1\times10^{-3}$  \\
				Quadratic Zeeman shift&$\phantom{0}1\times10^{-1}\pm2\times10^{-4}$&$\phantom{0\times10^{+0}}0\pm2\times10^{-4}$ \\
				AOM-induced phase chirp & $\phantom{0\times10^{+0}}0\pm O(1~\mathrm{Hz})$ & $\phantom{0\times10^{+0}}0\pm O(1~\mathrm{Hz})$ \\
				Linear Zeeman shift &$\phantom{0\times10^{+0}}0\pm3\times10^{2}\phantom{0}$&$\phantom{0\times10^{+0}}0\pm3\times10^{2}$ ($\sim1\times10^2$ after averaging)  \\
				Absolute frequency stability of PDH-locked probe laser &$\phantom{0\times10^{+0}}0\pm1\times10^{3}\phantom{0}$&$\phantom{0\times10^{+0}}0\pm1\times10^{3}$ ($\sim3\times10^2$ after averaging) \\
			\end{tabular}
		\end{ruledtabular}
		\label{tab:systematic_shifts}
	\end{table*}

	\section{Calculation of parameters associated with atomic wavefunctions}
	To evaluate quantities like $K_\alpha$, $F_\alpha$, $G_\alpha$, and $D_\alpha$ appearing in Eq.~1 in the main text, the change in the distribution of electrons in Yb$^+$ over space during the transitions of interest needs to be known. In particular, it is crucial to obtain $D_{\beta \alpha}$ to translate the measured nonlinearity into a new-boson-mediated neutron-electron coupling constant $y_e y_n$. Atomic structure calculations have been developed to numerically study electronic structures of atoms, and have been carried out here using some of the methods in the field as described in section \ref{section:calculation_methode}.
	
	\subsection{Description of methods}\label{section:calculation_methode}
	Here, the Dirac-Hartree-Fock (DHF) method \cite{Grant1980,Dyall1989} followed by the configurational interaction (CI) method \cite{Jonsson1996,Porsev2009,Fawcett1991,Biemont1998} has been used to calculate the two transitions in Yb$^{+}$. The calculation is relatively reliable because there is only one valence electron. Nevertheless, the full electron calculation is required to obtain the perturbed core-electron wavefunction due to the change in the valence-electron state; the change in core $s$ orbitals gives the major contribution to the sensitivity of $y_ey_n$ in the high-$m_\phi$ regime ($\gtrsim 1$~MeV). More advanced methods, for instance CI combined with many-body perterbation theory (CI+MBPT) \cite{Dzuba1996,Berengut2018,Dzuba2010} and CI+All-order method \cite{Safronova2009,Safronova2012}, have been developed. The calculation with the MBPT method (not combined with CI method) has been performed independently, and the results are compared in section~\ref{section:th_vs_exp_comparison}, as well as the main text, to provide an estimate of the systematic uncertainty of the calculation.
	
	The DHF and CI calculation were done with GRASP2018 \cite{FroeseFischer2018}. The DHF for closed core, from $1s$ to $5p$ subshells, was calculated first to obtain the basis set for core electrons. Then a valence electron was added and the basis set for all electrons was calculated. Finally, the correlation orbitals were added layer by layer to get better accuracy by taking core-core and core-valence correlation effects into account, and achieve convergence. CI calculations followed.
	
	Once the wavefunctions were obtained, the change in radial electron density functions $\rho_\alpha(r)$ \footnote{For he radial electron density $\rho_\alpha^{g,e}(r)$ for ground and excited states in transition $\alpha$, $\int_0^\infty \D r \rho_\alpha^{g,e}(r)=Z-I$ where $I$ is the charge number of ion and $\rho_\alpha(r) = \rho_\alpha^{e}(r) - \rho_\alpha^{g}(r)$} during transition $\alpha$ was calculated from the wavefunctions (it is not a part of GRASP2018), and all wavefunction-dependent quantities were obtained from the electron density $\rho_\alpha(r)$'s.
	
	The software package RIS4 \cite{Ekman2019} was used to calculate $K_\alpha$ from the outputs of GRASP2018. The details of the method by which this is achieved can be found in Ref.~\cite{Ekman2019}.
	
	The Seltzer moment expansion relates field shifts and the expansion of $\rho_\alpha(r)$ at the origin \cite{Seltzer1969,Blundell1987,Mikami2017}:
	\begin{align}
	\rho_\alpha(r) &= r^2 \left[ \xi_\alpha^{(0)} + \xi_\alpha^{(2)} r^2 + \cdots \right] \\
	\nu_{\alpha ji}^{\mathrm{FS}} &= \sum_{k=0}^{\infty} \underbrace{\frac{c\alpha'Z}{2\pi}\frac{\xi_\alpha^{(k)}}{(k+2)(k+3)}}_{F_\alpha^{(k)}}\delta\expval{r^{k+2}}\label{eq:Seltzer}
	\end{align}
	where $\alpha' \approx 1/137$ in Eq.~\ref{eq:Seltzer} is the fine-structure constant and $Z$ is the proton number. $F_\alpha \equiv F_\alpha^{(0)}$ and $G^{(4)}_\alpha \equiv F_\alpha^{(2)}$ in our notation. %(see section \ref{section:SM_correction} for $G^{(4)}_\alpha$.).
	$\xi_\alpha^{(0)}$ and $\xi_\alpha^{(2)}$ obtained by fitting $\rho_\alpha(r)$ with a power series at the origin can be converted into $F_\alpha$ and $G^{(4)}_\alpha$, respectively.
	
	The shift in transition frequency due to a new boson $\nu_{\alpha ji}^\phi = \Expval{a_{ji}V_{ne}(r)}_\alpha/h = \upsilon_{ne} D_\alpha a_{ji}$ (all quantities are defined in the main text) gives an expression for $D_\alpha$:
	\begin{equation}
	D_\alpha(m_\phi) = \frac{c}{2\pi} \int_0^\infty \D r \rho_\alpha(r) \frac{e^{-m_\phi r c/\hbar}}{r}.
	\end{equation}
	The numerical calculation of $D_\alpha$ for a given $\rho_\alpha(r)$ and $m_\phi$ is straightforward.
	
	The quadratic FS (QFS) $G^{(2)}_\alpha \drtsq$ captures the change in wavefunction itself due to the change in nuclear size, which is illustrated in the expression for the electronic factor:
	\begin{equation}\label{eq:QFS}
	G^{(2)}_\alpha = \frac{1}{2}\frac{\partial{F_\alpha}}{\partial{\expval{r^2}}} = \frac{c\alpha'Z}{24\pi}\frac{\partial \xi_\alpha^{(0)}}{\partial \mr{2}}
	\end{equation}
	$G^{(2)}_\alpha$ is given as the rate of change in electron density at the origin as nuclear size changes, and evaluating it requires repeated atomic structure calculations while gradually varying the nuclear size.
	
	The electronic factors were also calculated using an independent method that combines Brueckner MBPT and the random phase approximation (RPA) \cite{Dzuba1985} implemented in \textsc{amb}{\footnotesize i}\textsc{t} \cite{Kahl2019}. The MBPT correction accounting for core-valence correlations was calculated to second order in the residual Coulomb interaction. This was written in terms of a non-local potential $\hat\Sigma$, which was then added to the Dirac-Fock potential and solved self consistently to give ``Brueckner'' valence orbitals and energies. The Yukawa matrix elements for individual levels were calculated directly using the overlap of the orbitals and the Yukawa operator. The Yukawa potential also polarises the core, and this effect is included using the random-phase approximation. The $D_\alpha$ (or $D_\beta$) is obtained by subtracting the Yukawa matrix elements of the corresponding levels in the transition $\alpha$ (or $\beta$). To obtain $G^{(4)}_{\alpha(\beta)} = \partial \nu_{\alpha(\beta)}/\partial\mr{4}$, the calculation for transition $\alpha(\beta)$ is repeated while changing the nuclear charge distribution in such a way that $\mr{4}$ changes but $\mr{2}$ is constant. Similarly, to obtain $G^{(2)}_{\alpha(\beta)} = \frac{1}{2} \partial^2 \nu_{\alpha(\beta)}/(\partial\mr{2})^2$, the calculation is repeated while changing $\mr{2}$ but keeping $\mr{4}$ constant.
	
	\subsection{Comparison of calculations with experimental results}\label{section:th_vs_exp_comparison}
\begin{figure}[h!]
		\centering
		\includegraphics[width=\columnwidth,trim=10 0 45 0,clip]{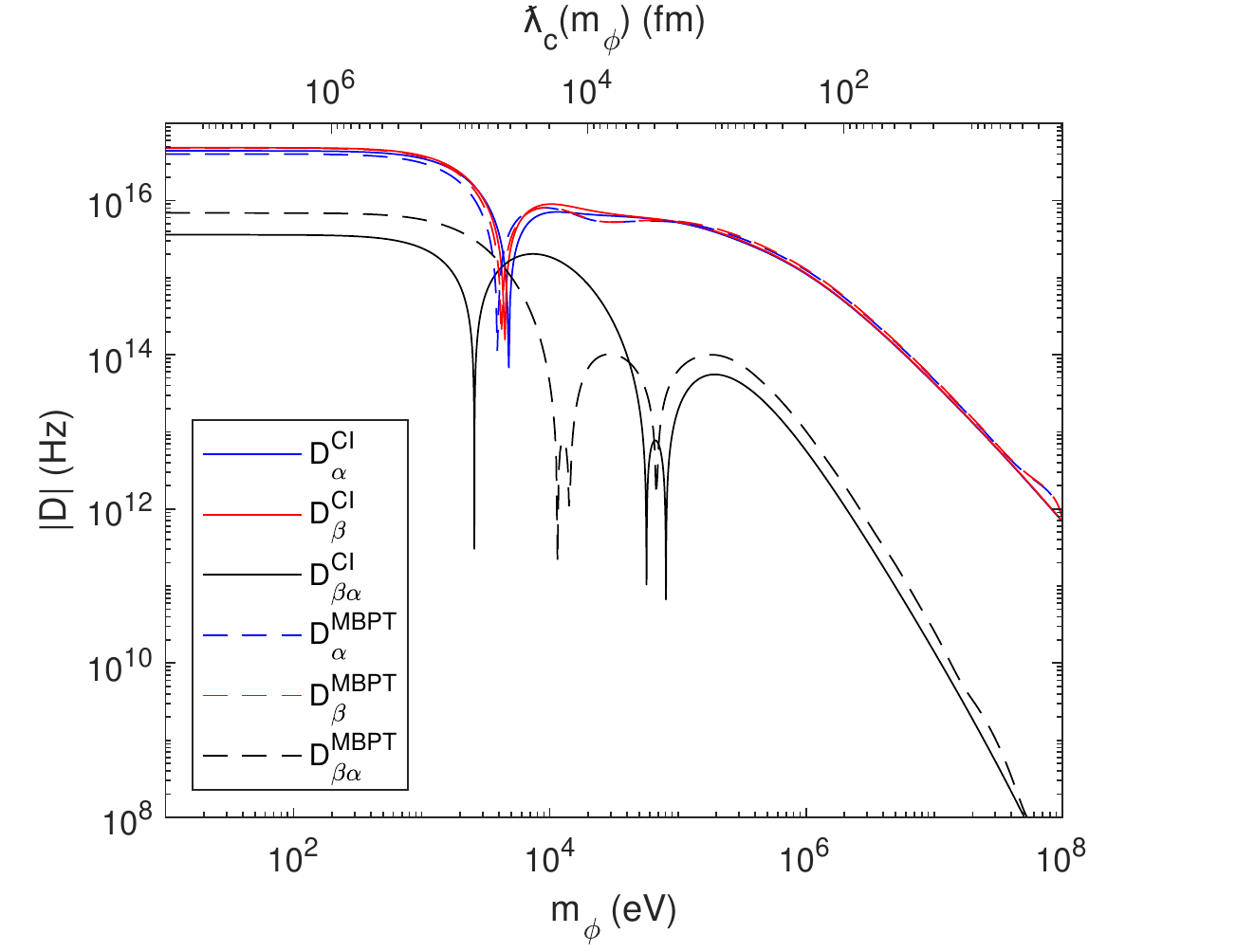}
		\caption{\label{fig:D_vs_m} Electronic factors associated with new-boson coupling $D_\alpha$, $D_\beta$, and $D_{\beta \alpha}$ vs. boson mass $m_\phi$ (bottom) and reduced Compton wavelength (top) for $\alpha=411$~ nm  and $\beta=436$~ nm transitions. The solid lines are for the CI calculation, and the dashed lines for the MBPT calculation. $D_\alpha$, $D_\beta$, and $D_{\beta\alpha}$ are positive at $m_\phi=1$~eV and change their signs at the peaks.}
	\end{figure}

	The accuracy of the calculated wavefunction using CI and MBPT methods is checked against our experimental data and literature values. The results are summarized in table~I, table~\ref{table:calculation_comparison}, table~\ref{table:ChargeRadius}, and Fig.~\ref{fig:D_vs_m}.
	
	\begin{table}[h!]
		\caption{\label{table:calculation_comparison}Quantities calculated using CI and MBPT methods, and estimated from the experiment for $\alpha$: 411 nm  and $\beta$: 436 nm transitions. $f_{\alpha,\beta} = \omega_{\alpha,\beta}/(2\pi)$ are the transition frequencies. Other quantities are defined in the main text.}
		\begin{ruledtabular}
			\begin{tabular}{ll|ccccccc}
				& & CI & MBPT & Exp. \\
				\hline
				$f_\alpha$ & [THz] & 808.11 & 764.86 & 729.47\footnotemark[1] \footnotemark[2]\\
				$f_\beta$  & [THz] & 770.13 & 717.94 & 688.36\footnotemark[1] \footnotemark[3] \\
				$F_\alpha$ & [GHz/fm$^2$] & -15.852 & -16.570 &  \\
				$F_\beta$  & [GHz/fm$^2$] & -16.094 & -16.771 &  \\
				$F_{\beta \alpha}$ & & 1.0153 & 1.0121 & 1.01141024(86) \\
				$K_\alpha$ & [GHz$\cdot$u] & -1678.3 & &  \\
				$K_\beta$  & [GHz$\cdot$u] & -1638.5 & &  \\
				$K_{\beta \alpha}$ & [GHz$\cdot$u] & 65 & & 120.208(23) \\
				$G^{(2)}_\alpha$ & [MHz/fm$^4$] & 40.23 & 43.47 &  \\
				$G^{(2)}_\beta$  & [MHz/fm$^4$] & 41.44 & 44.22 &  \\
				$G^{(2)}_{\beta \alpha}$ & [kHz/fm$^4$] & 232   & -36 & 59(17)\footnotemark[4] \\
				$G^{(4)}_\alpha$ & [MHz/fm$^4$] & 14.9291 & 10.338(3)\footnotemark[5] &  \\
				$G^{(4)}_\beta$  & [MHz/fm$^4$] & 15.1532 & 10.564(3)\footnotemark[5] &  \\
				$G^{(4)}_{\beta \alpha}$ & [kHz/fm$^4$] & -3.5 & -5(5)\footnotemark[5] & \\
				$D_\alpha$\footnotemark[6] & [THz] & 44145 & 40189 & \\
				$D_\beta$\footnotemark[6] & [THz]  & 48419 & 47962 & \\
				$D_{\beta\alpha}$\footnotemark[6]  & [THz] & 3602 & 6909 & 
			\end{tabular}
		\end{ruledtabular}
		\footnotetext[1]{The exact value varies by the few-GHz isotope shifts.}
		\footnotetext[2]{Ref.~\cite{Taylor1997,Roberts1999}}
		\footnotetext[3]{Ref.~\cite{Tamm2009,Webster2010}}
		\footnotetext[4]{If the observed nonlinearity comes purely from $G_{\beta \alpha}\drtsq$.}
		\footnotetext[5]{Numerical noise estimates}
		\footnotetext[6]{At $m_\phi = 1$~eV. Values over different $m_\phi$'s are shown in Fig.~\ref{fig:D_vs_m}.}
	\end{table}
	
	\subsubsection{Difference in the second nuclear charge moment $\dmr{2}$}
	
	The nuclear parameters $\lambda_{\alpha ji}$ \cite{Seltzer1969,Blundell1987,Fricke2004,Angeli2013}, which are essentially $\drt_{ji}$, can be obtained by dividing the measured FS, $F_\alpha \lambda_{\alpha ji}$, by the calculated $F_\alpha$. The FS $F_\alpha \lambda_{\alpha ji}$ can be obtained by subtracting $K_\alpha \mu_{ji}$ from the measured isotope shifts $\nu_{\alpha ji}$ ($K_\alpha$ is derived from the wavefunction calculation and $\mu_{ji}$ from mass spectroscopy \cite{AME2016_1,AME2016_2,Rana2012}). In short,
	\begin{equation}\label{eq:nuclear_param}
	\lambda_{\alpha ji} = \frac{\nu_{\alpha ji}^{\mathrm{FS}}}{F_\alpha} = \frac{\nu_{\alpha ji} - K_\alpha \mu_{ji}}{F_\alpha}.
	\end{equation}
	Plugging Eq.~\ref{eq:Seltzer} into Eq.~\ref{eq:nuclear_param} gives
	\begin{equation}\label{eq:lambda_vs_dmr2}
	\lambda_{\alpha ji} = \dmr{2}_{ji} \left[1 + \frac{G^{(4)}_\alpha}{F_\alpha} \frac{\dmr{4}_{ji}}{\dmr{2}_{ji}} + \cdots \right]
	\end{equation}
	To obtain $\dmr{2}$, the contribution of higher-order moments should be subtracted from $\lambda_{\alpha ji}$. Plugging Eq.~\ref{eq:dmr4_vs_dmr2} into Eq.~\ref{eq:lambda_vs_dmr2} gives
	% \begin{equation}\label{eq:lambda_vs_dmr2_2}
	\begin{align}
	\lambda_{\alpha ji} & = \dmr{2}_{ji} \left[ 1 + 2b \frac{G^{(4)}_\alpha \mr{2}_l}{F_\alpha} \left( 1 + \frac{\dmr{2}_{jl} + \dmr{2}_{il} }{2\mr{2}_l}\right ) \right] \label{eq:lambda_vs_dmr2_2} \\
	& \quad + O\left(\dmr{6}\right) \nonumber \\
	& \approx \dmr{2}_{ji} \left[ 1 + 2b \frac{G^{(4)}_\alpha \mr{2}_l}{F_\alpha}\right] \label{eq:lambda_vs_dmr2_3}
	\end{align}
	% \end{equation}
	where $l$ refers to some fixed isotope (here $l=172$). Using the values of $F_\alpha$ and $G^{(4)}_\alpha$ in table~\ref{table:calculation_comparison}, the value of $b$ for Yb in table~\ref{table:mr4_vs_mr2_test}, and $\mr{2}_l = 28.02(1)$~fm$^2$ from Ref.~\cite{Fricke2004}, the correction is calculated to be -6.97\%. The values from the CI calculation are used as they give the ratios $G^{(4)}_\alpha/F_\alpha = -9.418 \times 10^{-4}~\text{fm}^{-2}$ and $G^{(4)}_\beta/F_\beta = -9.416 \times 10^{-4}~\text{fm}^{-2}$,
	which are closer to the value given in a seminal paper about the Seltzer moment expansion: $-9.29 \times 10^{-4}~\text{fm}^{-2}$ \cite{Seltzer1969,Fricke2004} (this is approximately the universal across the levels and the transitions in a species of atom; see Ref.~\cite{Seltzer1969}.). The value of $b$ is assumed to vary less than about 1\% between the Yb isotopes (see table~\ref{table:mr4_vs_mr2_test}). The contribution of the term involving $\drt_{jl} + \drt_{il} \lesssim 0.4$~fm$^2$ in Eq.~\ref{eq:lambda_vs_dmr2_2} is $\lesssim 0.8\%$ of the -6.97\% correction from $\drt$. Therefore, the conversion from $\lambda_{\alpha ji}$ to $\drt$ via Eq.~\ref{eq:lambda_vs_dmr2_3} is precise to 0.1\%, unless the contribution of higher-order moments $\dmr{k>4}$ is larger than 0.1\%.
	
	The calculated mass shift, field shift, and $\dmr{2}$ values are shown in Table \ref{table:ChargeRadius}. The calculated $\dmr{2}$'s are in good agreement with the values in Ref. \cite{Angeli2013}, with differences of less than 10\%. $\dmr{2}$ values obtained by using  $F_\alpha$ and $F_\beta$ from the CI and MBPT calculations are compared in the table. For the analyses that use $\dmr{2}$ (e.g., the QFS line in Fig.~3), the values of $\dmr{2}$ that involve MBPT calculation are used, since the MBPT method is expected to better capture valence-core electron correlation. Note that, however, any difference in $F_\alpha$ or $F_\beta$ will change only the overall scale of $\dmr{2}$ values of different isotopes (see Eqs.~\ref{eq:nuclear_param} and \ref{eq:lambda_vs_dmr2_3}). Such differences would not change the pattern of nonlinearity originating from $\dmr{2}_{ji}$, or the direction of the QFS line in Fig.~3a.
	
	\begin{table*}
		\caption{\label{table:ChargeRadius}Mass shifts, field shifts and nuclear charge radius difference $\drt$ from measured isotope shifts and coefficients $K$, $F$ from atomic-structure calculation in table~\ref{table:calculation_comparison} for transitions $\alpha$: 411~nm and $\beta$: 436~nm.}
		\begin{ruledtabular}
			\begin{tabular}{cccccccccc}
				\multirow{2}{*}{Isotope pairs} & \multicolumn{2}{c}{\multirow{2}{*}{$K\mu_{ji}$ [MHz]}} & \multicolumn{2}{c}{\multirow{2}{*}{$F\lambda_{ji}$ [MHz]}} & \multicolumn{5}{c}{$\dmr{2}$ [fm$^2$]} \\
				& & & & & \multicolumn{2}{c}{CI} & \multicolumn{2}{c}{MBPT} & \\
				($j$,$i$) & $\alpha$ & $\beta$ & $\alpha$ & $\beta$ & $\alpha$ & $\beta$ & $\alpha$ & $\beta$ & Ref. \cite{Angeli2013}\footnote{Only statistical errors are presented in the parentheses. The large systematic errors in the electronic factors are not taken into account.} \\
				\hline 
				(168,170) & -117.7 & -114.9 & 2297 & 2327 & -0.156 & -0.155 & -0.149 & -0.149 & -0.1561(3) \\
				(170,172) & -115.0 & -112.2 & 2160 & 2189 & -0.146 & -0.146 & -0.140 & -0.140 & -0.1479(1) \\
				(172,174) & -112.4 & -109.7 & 1695 & 1719 & -0.115 & -0.115 & -0.110 & -0.110 & -0.1207(1) \\
				(174,176) & -109.9 & -107.3 & 1619 & 1641 & -0.110 & -0.110 & -0.105 & -0.105 & -0.1159(1)
			\end{tabular}
		\end{ruledtabular}
	\end{table*}
	
	\subsubsection{Slope and y-intercept in King plot}
	The slope and $y$-intercept $F_{\beta \alpha}$ and $K_{\beta \alpha}$ in the (standard) King plot can be obtained from the calculated $F_\alpha$, $F_\beta$, $K_\alpha$, and $K_\beta$, as defined in the main text. The calculated slopes $F_{\beta \alpha}^{\mathrm{CI}}$ and $F_{\beta \alpha}^{\mathrm{MBPT}}$ show excellent agreement with the experimental value of $F_{\beta \alpha}^{\mathrm{exp}}$, while the $y$-intercept $K_{\beta \alpha}^{\mathrm{CI}}$ is of the same order as the experimental value $K_{\beta \alpha}^{\mathrm{exp}}$ (see the main text or table~\ref{table:calculation_comparison}). Note, however, that the calculation of the mass shift coefficient is known to be a challenging task \cite{Papoulia2016,Puchalski2010}. We note that $K_{\beta \alpha}$ is the marginal remainder after $K_\beta$ and $K_\alpha$ for the two $D$ states with very similar wavefunctions are mostly cancelled out (at the level of 96\%). Therefore, the disagreement of the calculated $K_{\beta \alpha}$ to the experimental value does not necessarily imply the departure of the mass shift coefficients of each transition $K_\alpha$ or $K_\beta$ from their true values by the same factor.
	
	$F_\alpha$ is determined by the value of the wavefunctions at the origin (Eq.~\ref{eq:Seltzer}). Therefore, the good agreement of $\dmr{2}$ and $F_{\beta \alpha}$ with the experimental values implies that the calculated wavefunctions are reliable near the origin, the region which provides the dominant contribution to the sensitivity to $y_ey_n$ in the high-$m_\phi$ regime.
	
	\subsection{Estimation of nonlinearities within the SM}\label{section:SM_correction}
	The dominant SM contributions to the nonlinearity are expected to originate from higher-order FS terms,
	\begin{align}
	\label{SM_HO_shifts1}
	\delta \nu^{(2)}_\alpha &= G^{(2)}_\alpha \drtsq + G^{(4)}_\alpha \drf \\
	\label{SM_HO_shifts2}
	& = G_\alpha \drtsq
	\end{align}
	where $G_\alpha = G^{(2)}_\alpha + b G^{(4)}_\alpha$ is the effective QFS electronic factor (see Eq.~\ref{eq:mr4_vs_mr2} for $b$), as the higher-order terms in mass shift, $\alpha^2(m/M)^2$, are negligibly small \cite{Palmer1987}. The first term in Eq.~\ref{SM_HO_shifts1} (QFS) is from the second-order perturbation of the FS (Eq.~\ref{eq:QFS}), while the second term is the second leading-order moment in the Seltzer expansion for the first-order perturbation of the FS (Eq.~\ref{eq:Seltzer}). The correlation between $\mr{4}$ and $\mr{2}^2$ in Eq.~\ref{eq:mr4_vs_mr2} gives the next relation Eq.~\ref{SM_HO_shifts2} (see below in this section), and the two effects are combined with the same nuclear factor $\drtsq$. The contribution of this effective QFS to the nonlinearity is given as
	\begin{equation}
	\begin{aligned}
	G_{\beta\alpha}\drtsq &= \left[G_{\beta} - F_{\beta\alpha}G_{\alpha}\right] \drtsq \\ &= G^{(2)}_{\beta\alpha}\drtsq + b G^{(4)}_{\beta\alpha}\drtsq
	\label{eq:EffectiveqFS}
	\end{aligned}
	\end{equation}
	and the contributions from the QFS and the fourth-order Seltzer moment can be estimated separately.
	
	For the QFS, We calculate $G^{(2)}_\alpha = 40.23~\mathrm{MHz/fm^4}$ and $G^{(2)}_\beta = 41.44~\mathrm{MHz/fm^4}$ for the $\alpha= 411$~nm and $\beta=436$~nm transitions, respectively, using the CI calculation as in table~\ref{table:calculation_comparison}. For the fourth-order Seltzer moment, $G^{(4)}_\alpha$ and $G^{(4)}_\beta$ were calculated to be $14.9291~\mathrm{MHz/fm^4}$ and $15.1532~\mathrm{MHz/fm^4}$. The $F_\alpha$ and $G^{(4)}_\alpha$ are highly correlated (i.e., $F_\beta/F_\alpha \approx G^{(4)}_\beta/G^{(4)}_\alpha$), which suppresses $G^{(4)}_{\beta \alpha}$ by a factor of $\sim 2 \times 10^4$, giving $G^{(4)}_{\beta \alpha} = -3.5~\mathrm{kHz/fm^4}$, while the suppression for the QFS is $\sim 200$, yielding $G^{(2)}_{\beta\alpha} = 232~\mathrm{kHz/fm^4}$. The different suppression makes the contribution of QFS to the nonlinearity much bigger (by a factor of $\sim 66$) than that of fourth-order Seltzer moment although $G^{(4)}_\alpha$ is smaller than $G^{(2)}_\alpha$ only by a factor $\sim 2.7$.
	
	The nuclear factor $\drtsq \lesssim 0.07$~fm$^4$ gives $G_{\beta\alpha}\drtsq \lesssim 15$~kHz. The out-of-plane components of $\drtsq$ are $\lesssim 0.025$~fm$^4$ and thus the nonlinearity is the order of $\lesssim 5$~kHz.
	
	To justify Eq. \ref{SM_HO_shifts2} that absorbs the shape FS term $\dmr{4}$ into an effective QFS $(\dmr{2})^2$, we note that we expect the correlation
	\begin{equation}\label{eq:mr4_vs_mr2}
	\mr{4}_i = b \mr{2}_i^2
	\end{equation}
	to hold to a good approximation, where $b \approx 1$ is identical over different isotopes. This equation implies that the \textit{shape} of the charge distribution is preserved while the size varies between different isotopes: $\rho_{n,j}(r) = \rho_{n,i}(r/\epsilon_{ji})/\epsilon_{ji}$. Eq. \ref{eq:mr4_vs_mr2} is expected to hold for heavy ions in the absence of shell effects to order $1/A$ or better, where $A$ is the atomic mass. From Eq.~\ref{eq:mr4_vs_mr2}, one obtains the relation between $\dmr{4}$ and $\dmr{2}$ as follows:
	\begin{equation}
	\begin{aligned}
	\dmr{4}_{il} &= \mr{4}_i - \mr{4}_l \\
	&= b \left[ \mr{2}_i^2 - \mr{2}_l^2 \right] \\
	&= b \left[ \left(\mr{2}_l + \dmr{2}_{il}\right)^2 - \mr{2}_l^2 \right] \\
	&= b \left[ 2\mr{2}_l \dmr{2}_{il} + (\dmr{2}_{il})^2 \right]
	\end{aligned}
	\end{equation}
	where $\mr{n}_l$ is for a fixed reference isotope $l$. Consequently,
	\begin{equation}\label{eq:dmr4_vs_dmr2}
	\begin{aligned}
	\dmr{4}_{ji} &= \dmr{4}_{jl} - \dmr{4}_{il} \\
	&= b \left[2 \mr{2}_l \dmr{2}_{ji} + \left[\dmr{2}^2\right]_{ji} \right] \\
	&= 2 b \mr{2}_l \dmr{2}_{ji} \left[ 1 + \frac{\dmr{2}_{jl} + \dmr{2}_{il} }{2\mr{2}_l} \right]
	\end{aligned}
	\end{equation}
	where $[\dmr{2}^2]_{ji} \equiv (\dmr{2}_{jl})^2 - (\dmr{2}_{il})^2$.
	The last two rows show that $\dmr{4}$ is nearly linear in $\dmr{2}$ for small change in size ($\dmr{2}_{jl},~\dmr{2}_{il} \ll \mr{2}_l$), and the nonlinearity is due to $[\dmr{2}^2]$. The linear term $2b \mr{2}_l \dmr{2}_{ji}$ is absorbed into the leading-order FS $F_\alpha\mr{2}$, while the nonlinear term is combined with the QFS: $(G^{(2)}_\alpha + b G^{(4)}_\alpha) [\dmr{2}^2]_{ji}$. 
	
	The assumption Eq.~\ref{eq:mr4_vs_mr2} can be tested using the nuclear charge distribution $\rho_n(r)$ measured by electron scattering experiments \cite{DeVries1987,Sasanuma1979}. The Fourier-Bessel coefficients in Ref.~\cite{DeVries1987} are used to retrieve $\rho_n(r)$ of each isotope and $\mr{2}$, $\mr{4}$, and $b$ are calculated using the $\rho_n(r)$. The results for Pd, Sm, and Pb are listed in table~\ref{table:mr4_vs_mr2_test}. The variations of $b$ over different isotopes are indeed small: 0.1 to 1\%. The difference may be merely due to the experimental uncertainly of the data in Ref.~\cite{DeVries1987}.
	
	\begin{table}
		\caption{\label{table:mr4_vs_mr2_test} $\mr{4}$-to-$\mr{2}^2$ ratio $b$ (in Eq.~\ref{eq:mr4_vs_mr2}) for the isotopes of Pd, Sm, and Pb calculated using electron scattering data \cite{DeVries1987}. The fractional variation $\Delta b/\overline{b}$, the standard deviation of the $b$ coefficients divided by their mean, is shown in the last column for each element. $b$ of $^{174}$Yb from the electron scattering data in \cite{Sasanuma1979} is shown for comparison.}
		\begin{ruledtabular}
			\begin{tabular}{ccccc}
				Element & Isotope & $b$ & $\Delta b/\overline{b}$ \\
				\hline
				\multirow{4}{*}{Pd} & 104 & 1.3173 & \multirow{4}{*}{0.0026} \\
				& 106 & 1.3225 \\
				& 108 & 1.3247 \\
				& 110 & 1.3246 \\
				\hline
				\multirow{5}{*}{Sm} & 144 & 1.2861 & \multirow{5}{*}{0.0129} \\
				& 148 & 1.2979 & \\
				& 150 & 1.3175 & \\
				& 152 & 1.3212 & \\
				& 154 & 1.3254 & \\
				\hline
				\multirow{4}{*}{Pb} & 204 & 1.2755 & \multirow{4}{*}{0.0011}\\
				& 206 & 1.2735 & \\
				& 207 & 1.2770 & \\
				& 208 & 1.2758 & \\
				\hline
				Yb & 174 & 1.3211 &
			\end{tabular}
		\end{ruledtabular}
	\end{table}
	
	Finally, note that the choice of $l$ in the definition of $\drtsq_{ji}$ does not change the associated nonlinearity, as shown below:
	\begin{align}
	\drtsq_{ji}^l &\equiv (\drt_{jl})^2 - (\drt_{il})^2 \nonumber \\
	& = (\drt_{jm} + \drt_{ml})^2 - (\drt_{im} + \drt_{ml})^2 \nonumber \\
	& = \drtsq_{ji}^m + 2\drt_{ml}\drt_{ji} \label{eq:drtsq_l_vs_m}
	\end{align}
	where $l$, $m$ are some fixed isotopes. The last term in Eq.~\ref{eq:drtsq_l_vs_m} is proportional to $\drt_{ji}$ and is readily absorbed into the field shift term.

	In summary, the QFS $G^{(2)}_{\beta \alpha} \drtsq$ gives the dominant contribution to the SM nonlinearity unless the shape of nuclear charge distribution varies strongly between different isotopes. The contributions from other nuclear effects that are not considered here are smaller, e.g., the nuclear polarizability that is expected to contribute at the level $<10~\mathrm{Hz})$ \cite{Flambaum2018}.
	
	\section{Generalization of nonlinearity measures to more isotopes and transitions}
	The nonlinearity measures $\zeta_\pm$ can be generalized for more than 4 pairs of isotopes or more than two transitions. For $N$ pairs of isotopes (i.e. $N+1$ isotopes), $N-2$ independent parameters characterizing the pattern of nonlinearity can be defined, and visualized in an $(N-2)$-dimensional plot. This can be easily understood with a vectorial approach. Let $\vec{x} = (x_{j_1i_1},\cdots,x_{j_Ni_N})$ where $x_{j_qi_q}$ is a term in Eq.~3 (e.g., $G_{\beta \alpha} \overline{\drtsq}_{j_qi_q}$) for the $q$-th isotope pair. After the components along the \textit{King vectors} $\vec{\overline{\mu}} = (1,\cdots,1)$ (mass shift) and $\vec{\overline{\nu}}_\alpha = (\overline{\nu}_{\alpha j_1i_1},\cdots,\overline{\nu}_{\alpha j_Ni_N})$ (field shift) are subtracted from $\vec{x}$ via linear fitting, the remaining vector can be uniquely determined by the components of $(N-2)$ vectors $\{\vec{\zeta}_1,\cdots,\vec{\zeta}_{N-2}\}$ which are linearly independent from each other and from $\{\vec{\overline{\mu}}, \vec{\overline{\nu}}_\alpha\}$.
	
	A generalization for more than two transitions can be obtained by pairing transitions; $M-1$ independent pairs of transitions can be formed out of $M$ transitions and the analysis with a two-dimensional King plot described throughout the paper can be applied to each pair. The nonlinearity measures $\zeta_\pm$ (or $(\zeta_1,\cdots,\zeta_{N-2})$ for the case of $N+1$ isotopes) from each pair of transitions can be displayed in the same two-dimensional (or $(N-2)$-dimensional) plot. In particular, if all of the transition pairs share a certain transition $\alpha$ (i.e., the pairs are $(\alpha,\beta),(\alpha,\gamma),(\alpha,\delta),\cdots$), the King vectors $\{\vec{\overline{\mu}}, \vec{\overline{\nu}}_\alpha\}$ are the same for all pairs of transitions and comparing the nonlinearities becomes straightforward.
	
% 	\nocite{CetinaThesis2011,Olmschenk2007,Martensson-Pendrill1994,Feldker2018,Sugiyama2000,Gerz1988,Lizuain2007,Berkeland1998,Safronova2012,Roy2017,Nandy2014,Dube2013,Degenhardt2005,Sanner2018,TrappedChargedParticlesBook,Dzuba1996,Dzuba2010,Safronova2009,Ekman2019,Blundell1987,Roberts1999,Tamm2009,Webster2010,Fricke2004,Papoulia2016,Puchalski2010,Palmer1987,DeVries1987,Sasanuma1979}
	
%	\bibliography{IsotopeShiftPaperRefs}
%	% \bibliographystyle{bst/apsrev4-0_edited}

\providecommand{\noopsort}[1]{}\providecommand{\singleletter}[1]{#1}%